\documentclass[11pt]{article}

\usepackage[review]{acl}

\usepackage{times}
\usepackage{latexsym}
\usepackage{amsmath,amssymb}
\usepackage{tabularx}
\usepackage[T1]{fontenc}
\usepackage{enumitem}
\usepackage{url}
\usepackage[utf8]{inputenc}
\usepackage{titletoc}
\usepackage{cleveref}
\crefname{appendix}{}{ }
\usepackage{microtype}

\usepackage{inconsolata}
\usepackage{xcolor}
\usepackage{tcolorbox}
\tcbuselibrary{listings,breakable,skins}
\usepackage{listings}

\usepackage{graphicx}
\usepackage{booktabs}   
\usepackage{multirow}   
\usepackage{array}      
\usepackage[table]{xcolor} 
\definecolor{Gray2}{gray}{0.5}
\definecolor{Gray3}{gray}{0.5}
\definecolor{RowHighlight}{gray}{0.9}
\title{You Only Anonymize What Is Not Intent-Relevant:\\
Suppressing Non-Intent Privacy Evidence}

\author{
 \textbf{Weihao Shen\textsuperscript{1}},
 \textbf{Yaxin Xu\textsuperscript{2}},
 \textbf{Shuang Li\textsuperscript{1}},
 \textbf{Wei Chen\textsuperscript{1}},
\\
 \textbf{Yuqin Lan\textsuperscript{1}},
 \textbf{Meng Yuan\textsuperscript{1}},
 \textbf{Fuzhen Zhuang\textsuperscript{1}}
\\
 \textsuperscript{1}Institute of Artificial Intelligence, Beihang University, Beijing, China  \\
 \textsuperscript{2}State Key Laboratory of Information Engineering in Surveying, Mapping,\\ and Remote Sensing, Wuhan
University, Wuhan, China
\\
 \small{
   \{shenweihao, shuangliai, chenwei23, lanyq, yuanmeng97, zhuangfuzhen\}@buaa.edu.cn
 }\\
 \small{
   xuyaxin@whu.edu.cn
 }
}

\begin{document}
\nolinenumbers
\maketitle
\begin{abstract}
Anonymizing sensitive information in user text is essential for privacy, yet existing methods often apply uniform treatment across attributes, which can conflict with communicative intent and obscure necessary information.
This is particularly problematic when personal attributes are integral to expressive or pragmatic goals.
The central challenge lies in determining which attributes to protect, and to what extent, while preserving semantic and pragmatic functions.
We propose \textsc{IntentAnony}, a utility-preserving anonymization approach that performs intent-conditioned exposure control.
\textsc{IntentAnony} models pragmatic intent and constructs privacy inference evidence chains to capture how distributed cues support attribute inference.
Conditioned on intent, it assigns each attribute an exposure budget and selectively suppresses non-intent inference pathways while preserving intent-relevant content, semantic structure, affective nuance, and interactional function.
We evaluate \textsc{IntentAnony} using privacy inference success rates, text utility metrics, and human evaluation.
The results show an approximately 30\% improvement in the overall privacy--utility trade-off, with notably stronger usability of anonymized text compared to prior state-of-the-art methods. Our code is available at \url{https://github.com/Nevaeh7/IntentAnony}.

\end{abstract}
\begin{figure}[!t]
\setlength{\belowcaptionskip}{-4pt} 
\centering
 \includegraphics[width=1.0\linewidth]{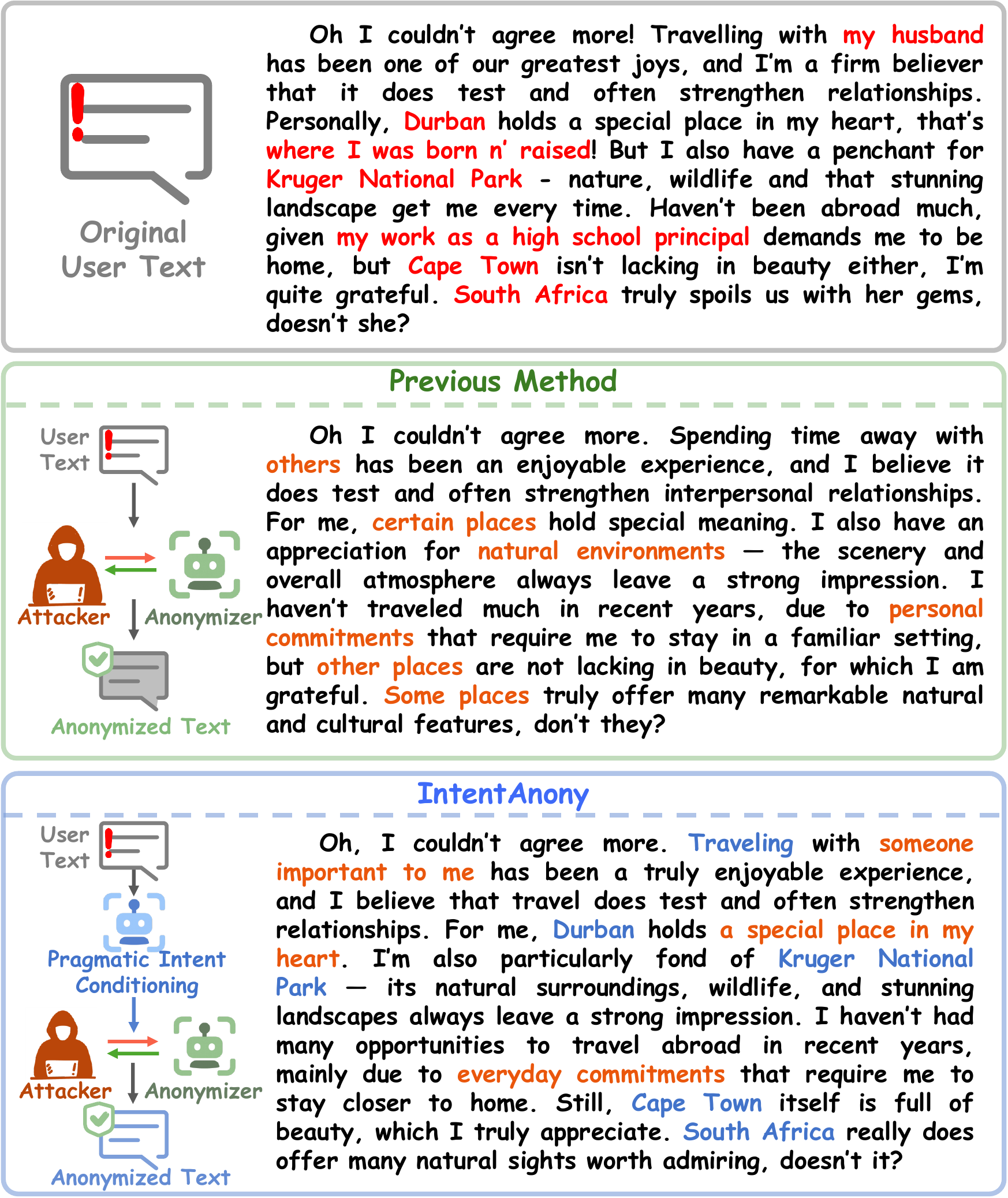}
 \caption{Anonymization under inference-based privacy threats.
Compared with uniform rewriting methods that obscure sensitive details at the cost of communicative intent, \textsc{IntentAnony} conditions anonymization on pragmatic intent and selectively suppresses non-intent privacy evidence, preserving semantic meaning and interactional function while reducing inference risk.
 }
 \vspace{-12pt}
\label{fig:introduction_examples}
\end{figure}
\section{Introduction}

Text anonymization aims to mitigate privacy risks in natural-language content while preserving the information necessary for faithful interpretation and effective use\cite{anony,datasec}.
This goal has become increasingly difficult in the era of large language models (LLMs), whose reasoning capabilities enable sensitive attribute inference even when explicit identifiers are removed\cite{mvsm, llmsecsurvey,wei2024privacy,tokenprivacy}.
Recent work shows that privacy leakage often arises from distributed semantic, stylistic, and contextual cues rather than surface-level memorization\cite{evalsec}, rendering conventional masking-based anonymization insufficient in many settings \cite{staab2023beyond, sarkar2024deidentification}.

These inference-based privacy threats~\cite{piinfer, mmprivacy, pleak} have motivated a shift from simple de-identification toward semantics-aware anonymization \cite{uia, du2025beyond}.
Recent approaches formulate anonymization as a constrained transformation problem that balances privacy protection and utility preservation through optimization, adversarial evaluation, or randomized rewriting strategies \cite{yang2025robust, frikha2024incognitext, kim2025seal}.
However, most existing methods apply privacy treatment in an attribute-agnostic manner, either uniformly masking detected attributes or broadly rewriting text without accounting for the communicative role that attributes play.

In practice, personal attributes vary in privacy sensitivity\cite{privacyci,pinvestigation}. Some are deliberately disclosed to support communicative goals, while others seem benign in isolation but collectively enable inference. Uniform anonymization thus faces a trade-off: overly aggressive suppression harms intent-related utility, whereas insufficient suppression leaves inference channels exploitable by capable attackers Figure~\ref{fig:introduction_examples}.

To address this limitation, we propose \textsc{IntentAnony}, an anonymization approach that formulates privacy protection as an intent-conditioned exposure control problem.
The central insight is that inference-based privacy risks emerge not from individual attributes alone, but from how distributed linguistic cues are pragmatically organized to support attribute inference under a given communicative goal.
Accordingly, \textsc{IntentAnony} explicitly models pragmatic intent and organizes privacy-relevant cues into privacy inference evidence chains, capturing how multiple textual spans jointly contribute to sensitive attribute inference.
Conditioned on the recognized intent, each attribute is assigned an allowable exposure budget through an intent--attribute exposure matrix, which regulates the amount of inferential support retained in the anonymized text.
This design enables selective suppression of non-intent inference pathways while preserving evidence that is functionally necessary for conveying meaning, stance, and interactional purpose.
By intervening directly on inference-supporting evidence structures rather than surface identifiers, \textsc{IntentAnony} reduces inference risk while maintaining semantic coherence and pragmatic intent.

We evaluate \textsc{IntentAnony} using automatic privacy--utility metrics and human assessment, showing consistent reductions in inference risk together with improved preservation of semantic content, emotional tone, and communicative intent compared to strong masking and LLM-based rewriting baselines.

The main contributions of this work are summarized as follows:
\begin{itemize}[leftmargin=1.0em, itemsep=0em]
    \item We introduce an intent-conditioned text anonymization approach that aligns privacy protection with communicative function by regulating personal attribute exposure.
    \item We propose a scene and intent conditioned exposure governance mechanism that enforces explicit attribute-level budgets for fine-grained, context-aware anonymization.
    \item Extensive privacy and utility evaluations, complemented by human evaluation, show that the proposed approach achieves a stronger privacy--utility trade-off, reducing attribute inference risk while preserving semantic coherence and communicative intent.
\end{itemize}

\section{Related Work}

\noindent \textbf{Text Anonymization.}
Recent work has moved text anonymization beyond simple PII masking~\cite{azure} toward LLM-aware approaches that explicitly account for inference-based privacy risks.
Many methods formulate anonymization as a constrained transformation problem that jointly optimizes privacy and utility, including iterative LLM-based frameworks with multi-component evaluation~\cite{yang2025robust} and their analyses in personalized writing settings~\cite{pasch2025balancing,manzanares2025comparative}.
Complementary directions include conditional anonymization via private attribute randomization~\cite{frikha2024incognitext}, self-refining anonymizers based on adversarial distillation~\cite{kim2025seal}, and adversarial frameworks that leverage LLM inference itself for anonymization~\cite{staab2025advanced}.
Additional work explores inference-aware sanitization~\cite{pilan2025truthful}, stylometric obfuscation~\cite{adversarialstylometry}, context-preserving anonymization for structured or domain-specific text~\cite{zarski2025contextprivacy}, and data-level anonymization for privacy-aware LLM deployment~\cite{gardiner2024dataanonym}.
Most existing methods are attribute-agnostic and overlook the communicative and pragmatic roles of personal attributes, which often results in unnecessary utility loss or insufficient disruption of inference pathways.

\noindent \textbf{Privacy and Utility Trade-offs.}
Balancing privacy protection and textual utility is particularly challenging in the presence of inference-capable LLM adversaries.
Prior work shows that modern LLMs can infer sensitive attributes from anonymized text even after explicit identifiers are removed~\cite{staab2023beyond}.
This has motivated research on privacy--utility trade-offs in LLM-based anonymization.
Recent methods suppress attribute inference through optimization- and evaluation-based strategies while preserving utility~\cite{yang2025robust,frikha2024incognitext}, and through inference-aware sanitization techniques~\cite{pilan2025truthful,manzanares2025comparative}.
Parallel studies examine broader privacy risks and defenses for LLMs, including inference attacks and mitigation via differential privacy~\cite{dpllm, sldp} or privacy-preserving inference~\cite{miranda2024preserving,yan2025llmprivacy}.
However, most existing methods lack intent-aware exposure control; in contrast, our work selectively suppresses non-intent inference cues while preserving intent-critical content.


\begin{figure*}[!t]
  \centering
  \includegraphics[width=1.00\textwidth]{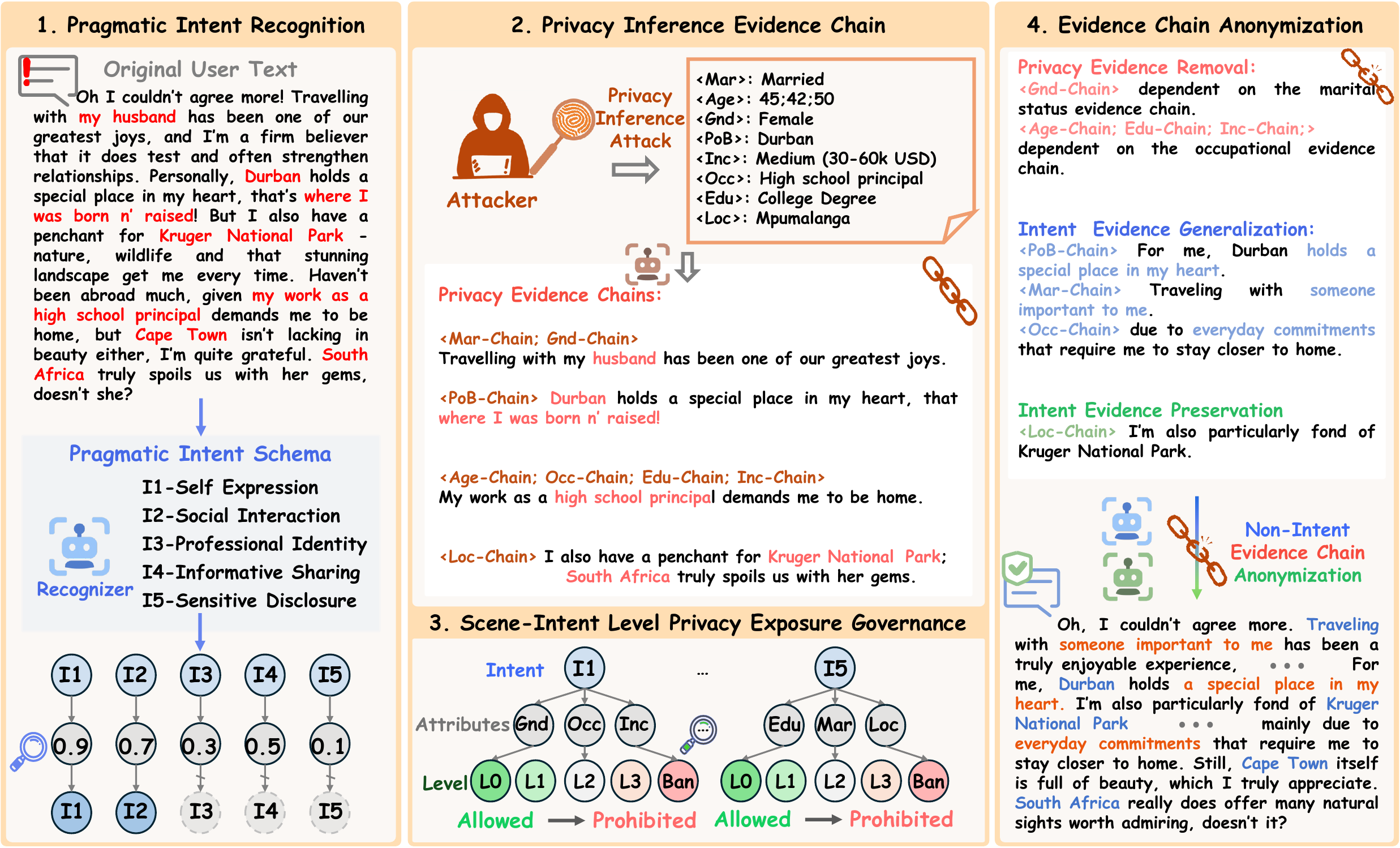}
\caption{Overview of the proposed intent-aware anonymization framework. The pipeline includes four stages: (1) pragmatic intent recognition to identify communicative intents in the input text; (2) privacy inference evidence chain construction that organizes sensitive attributes into intent-grounded evidence chains; (3) scene--intent level privacy exposure governance for determining appropriate anonymization levels; and (4) evidence chain anonymization, which selectively rewrites or removes non-intent evidence while preserving intent-relevant content.}

  \label{fig:prompting_arc}
\vspace{-10pt}
\end{figure*}

\section{Method}

We present \textsc{IntentAnony}, an intent-conditioned text anonymization approach that regulates personal attribute exposure according to communicative intent.
Privacy leakage often arises from distributed linguistic cues whose inferential effect depends on the pragmatic role of attributes, causing uniform anonymization to either undercut inference resistance or unnecessarily distort meaning.
\textsc{IntentAnony} addresses this by formulating anonymization as intent-conditioned exposure control, explicitly aligning privacy protection with communicative function.
Figure~\ref{fig:prompting_arc} provides an overview of the approach.

\subsection{Pragmatic Intent Recognition}

Given an input text $x$, the first step is to infer its underlying communicative intents.
Instead of task-oriented intents commonly used in dialogue systems, we focus on pragmatic intents that characterize how language is functionally employed in social and expressive contexts, such as self-expression, social interaction, identity presentation, informational sharing, and sensitive disclosure.
These intents govern not only what information is conveyed, but also which personal attributes are pragmatically relevant to meaning construction.

Formally, we define a finite set of pragmatic intents
\begin{equation}
\mathcal{I} = \{ I_1, I_2, \dots, I_K \},
\end{equation}
and cast intent recognition as a multi-label inference problem
\begin{equation}
f_{\text{intent}} : x \rightarrow \mathcal{I}(x),
\end{equation}
where $\mathcal{I}(x) \subseteq \mathcal{I}$ denotes the set of intents expressed in $x$.
Multiple intents may co-occur within a single text, reflecting the compositional nature of human communication.

We employ large language models as intent recognizers, exploiting their discourse-level reasoning to infer pragmatic intent from semantic and contextual evidence.
The resulting intent set acts as a global constraint that guides subsequent anonymization decisions by distinguishing intent-relevant information from privacy-risk evidence.

\subsection{Privacy Inference Evidence Chain}

Sensitive personal attributes are rarely disclosed through isolated identifiers.
Instead, they are often inferred from the aggregation of explicit, implicit, and contextual cues distributed across a text.
Such compositional inference enables the recovery of latent attributes even in the absence of direct mentions, which limits the effectiveness of surface-level anonymization.

To capture this structure, we introduce privacy inference evidence chains.
For each sensitive attribute $a \in \mathcal{A}$, an evidence chain is defined as
\begin{equation}
C_a = \{ e_{a,1}, e_{a,2}, \dots, e_{a,n} \},
\end{equation}
where each element $e_{a,i}$ denotes a textual cue contributing to the inference of $a$.
These cues may be lexical, semantic, or contextual, and their inferential strength arises from their joint configuration rather than from any single element.

Each evidence element is assessed with respect to the recognized pragmatic intents of the text, enabling a distinction between intent-relevant evidence that supports communicative function and non-intent evidence that primarily amplifies inference pathways.
Modeling privacy leakage at the level of evidence chains enables anonymization to intervene directly on inference-supporting structures, rather than relying on token-level masking or unstructured heuristic rewriting.

\subsection{Scene--Intent Level Privacy Exposure Governance}

To balance privacy protection and textual utility in an interpretable manner, we introduce an exposure governance mechanism conditioned on scene context and communicative intent.
Rather than relying on static or attribute-agnostic anonymization rules, the mechanism determines how much information about each attribute may be retained according to the communicative intent expressed in the text and its surrounding context.

We define an ordered set of privacy exposure granularity levels
\begin{equation}
\mathcal{E} = \{ L_0, L_1, L_2, L_3, \text{BAN} \},
\end{equation}
where increasing levels impose stricter exposure constraints, ranging from minimal modification to complete suppression.

Given a scene representation $s(x)$ inferred from context, we define an exposure governance function
\begin{equation}
G : \mathcal{S} \times \mathcal{I} \times \mathcal{A} \rightarrow \mathcal{E},
\end{equation}
which assigns each combination of scene, intent, and attribute a maximum allowable exposure level.
This mapping reflects how the appropriateness of attribute disclosure varies across different communicative contexts.

When a text expresses multiple intents, we adopt a conservative aggregation strategy and assign each attribute $a$ an effective exposure budget
\begin{equation}
\ell_a = \min_{I_k \in \mathcal{I}(x)} G\bigl(s(x), I_k, a\bigr).
\end{equation}
The resulting budget constrains the total contribution of the corresponding evidence chain $C_a$ in the anonymized output.
By conditioning exposure control on both scene context and communicative intent, this mechanism enables fine-grained anonymization while avoiding insufficient protection and unnecessary removal of intent-relevant information.

\subsection{Evidence Chain Anonymization}

Given the intent-conditioned exposure budgets, \textsc{IntentAnony} generates an anonymized text $\tilde{x}$ through evidence-chain-guided rewriting.
Anonymization operates at the level of privacy inference evidence chains rather than isolated tokens, enabling direct intervention on the structured aggregation of cues that supports attribute inference.

For each sensitive attribute $a \in \mathcal{A}$ with evidence chain $C_a$, we assess the functional role of individual evidence elements with respect to the recognized communicative intents $\mathcal{I}(x)$.
Based on this assessment, the evidence chain $C_a$ is conceptually decomposed into two complementary subsets: an intent-relevant component $C_a^{\text{intent}}$, which is necessary for realizing the communicative intent, and a non-intent component $C_a^{\text{non-intent}}$, which primarily strengthens sensitive attribute inference without contributing to the intended meaning.
This decomposition allows intent-relevant evidence to be preserved or generalized, while non-intent evidence is selectively attenuated to reduce inference-based privacy risk.

Anonymization is governed by the exposure budget $\ell_a$ associated with attribute $a$, which constrains the expected inference risk induced by retained evidence.
Specifically, the anonymized text $\tilde{x}$ is required to satisfy
\begin{equation}
\mathbb{E}\big[\, \mathcal{R}(a \mid \tilde{x}) \,\big] \;\leq\; \ell_a,
\end{equation}
where $\mathcal{R}(a \mid \tilde{x})$ denotes the inference risk of attribute $a$ given the text, as estimated by privacy-oriented inference prompts, and the expectation is taken over the stochastic generation process of the language model.
To meet this constraint, evidence in $C_a^{\text{non-intent}}$ is suppressed or removed, while evidence in $C_a^{\text{intent}}$ is preserved through abstraction or generalization when necessary.
This process is implemented via constrained LLM-based rewriting, in which prompts explicitly encode intent requirements together with attribute-specific exposure limits.
By intervening at the level of evidence chains and explicitly constraining inference risk, \textsc{IntentAnony} disrupts compositional inference pathways while preserving semantic coherence and interactional quality.

\section{Experimental Set-up}
\begin{table*}[t]
    \centering
    \footnotesize
    \renewcommand{\arraystretch}{0.85}
    \setlength{\tabcolsep}{5pt}
    \begin{tabular}{l*{12}{c}}
        \toprule[1pt]

        \multirow{3}{*}{\textbf{Metric}} 
        & \multicolumn{6}{c}{\textbf{PersonalReddit}} 
        & \multicolumn{6}{c}{\textbf{SynthPAI}} \\
        \cmidrule(lr){2-7} \cmidrule(lr){8-13}

        & \textbf{Orig.} & \textbf{Azure} & \textbf{Dipper} & \textbf{A.A.} & \textbf{RUPTA} & \textbf{Ours}
        & \textbf{Orig.} & \textbf{Azure} & \textbf{Dipper} & \textbf{A.A.} & \textbf{RUPTA} & \textbf{Ours} \\

        \midrule[0.25pt]
        \hspace{1em} Age
        & 0.500 & 0.379 & 0.482 & 0.329 & 0.379 & 0.400
        & 0.387 & 0.342 & 0.385 & 0.325 & 0.333 & 0.333 \\

        \hspace{1em} Edu
        & 0.765 & 0.643 & 0.724 & 0.520 & 0.735 & 0.541
        & 0.690 & 0.563 & 0.688 & 0.344 & 0.469 & 0.375 \\

        \hspace{1em} Gnd
        & 0.890 & 0.787 & 0.856 & 0.748 & 0.780 & 0.732
        & 0.943 & 0.829 & 0.854 & 0.756 & 0.707 & 0.780 \\

        \hspace{1em} Inc
        & 0.736 & 0.755 & 0.712 & 0.679 & 0.651 & 0.745
        & 0.745 & 0.658 & 0.632 & 0.539 & 0.632 & 0.605 \\

        \hspace{1em} Loc
        & 0.593 & 0.102 & 0.561 & 0.163 & 0.045 & 0.098
        & 0.328 & 0.103 & 0.362 & 0.000 & 0.138 & 0.138 \\

        \hspace{1em} Mar
        & 0.757 & 0.486 & 0.806 & 0.446 & 0.365 & 0.338
        & 0.833 & 0.660 & 0.766 & 0.426 & 0.596 & 0.489 \\

        \hspace{1em} Occ
        & 0.574 & 0.347 & 0.512 & 0.231 & 0.546 & 0.204
        & 0.707 & 0.634 & 0.704 & 0.127 & 0.634 & 0.338 \\

        \hspace{1em} PoB
        & 0.549 & 0.155 & 0.451 & 0.099 & 0.014 & 0.099
        & 0.333 & 0.000 & 0.429 & 0.000 & 0.143 & 0.000 \\

        \midrule[0.5pt]
        \rowcolor{RowHighlight} \textbf{Privacy $\downarrow$}
        & 0.650 & 0.411 & 0.614 & \underline{0.365} & 0.417 & \textbf{0.353}
        & 0.607 & 0.499 & 0.579 & \textbf{0.334} & 0.474 & \underline{0.410} \\

        \midrule[0.25pt]
        \hspace{1em} Mean
        & 1.000 & 0.787 & 0.937 & 0.833 & 0.826 & 0.956
        & 1.000 & 0.727 & 0.854 & 0.847 & 0.920 & 0.981 \\

        \hspace{1em} Read
        & 1.000 & 0.371 & 0.970 & 0.998 & 0.990 & 1.000
        & 1.000 & 0.291 & 0.827 & 0.988 & 0.862 & 0.946 \\

        \hspace{1em} Hall
        & 1.000 & 1.000 & 0.948 & 0.991 & 0.873 & 1.000
        & 1.000 & 1.000 & 0.834 & 0.995 & 0.961 & 0.998 \\

        \midrule[0.25pt]
        \hspace{1em} BLEU
        & 1.000 & 0.824 & 0.247 & 0.622 & 0.745 & 0.852
        & 1.000 & 0.798 & 0.130 & 0.490 & 0.712 & 0.849 \\

        \hspace{1em} ROUGE
        & 1.000 & 0.957 & 0.676 & 0.804 & 0.878 & 0.930
        & 1.000 & 0.952 & 0.617 & 0.730 & 0.912 & 0.944 \\

        \midrule[0.5pt]
        \rowcolor{RowHighlight} \textbf{Utility $\uparrow$}
        & 1.000 & 0.833 & 0.625 & 0.789 & \underline{0.840} & \textbf{0.923}
        & 1.000 & 0.807 & 0.528 & 0.721 & \underline{0.846} & \textbf{0.923} \\

        \midrule[0.75pt]
        \rowcolor{RowHighlight} \textbf{Overall $\uparrow$}
        & - & 0.201 & -0.320 & \underline{0.227} & 0.198 & \textbf{0.379}
        & - & -0.015 & -0.426 & \underline{0.171} & 0.065 & \textbf{0.247} \\

        \bottomrule[1pt]
    \end{tabular}
\caption{
Unified performance comparison of anonymization methods on the PersonalReddit and SynthPAI datasets.
A.A.\ denotes Adv.\ Anon.; best and second-best results are shown in bold and underlined, respectively.
}
    \label{table:combined-datasets}
    \vspace{-0.15in}
\end{table*}

\noindent \textbf{Datasets.}
We evaluate our method on two datasets covering realistic and controlled settings.
Both datasets consist of Reddit-style text annotated with personal attributes and exhibit properties comparable to authentic user-generated content.
\textit{PersonalReddit}\cite{staab2023beyond} contains Reddit-style Q\&A pairs with naturally embedded attributes, while \textit{SynthPAI}~\cite{synthpai} consists of Reddit-style comments with systematically varied attribute combinations.
Additional data preprocessing details are provided in Appendix~\ref{sec:data_impl}.

\noindent \textbf{Evaluation Metrics.}
Utility is evaluated along four complementary dimensions: readability, hallucination control, semantic preservation, and surface-level similarity.
Readability and hallucination capture textual fluency and faithfulness, while semantic preservation assesses retention of the original meaning and intent; BLEU and ROUGE quantify lexical overlap.
Privacy protection is evaluated via attribute inference attacks by a strong LLM adversary, where attack accuracy indicates residual privacy leakage.
We jointly report privacy and utility to assess whether anonymization reduces inference risk while preserving communicative value.

\medskip
\noindent \textbf{Models in Comparison.}
We compare \textsc{IntentAnony} with representative anonymization systems covering commercial solutions, rule-based pipelines, and LLM-driven approaches.
\begin{itemize}[leftmargin=1.0em, itemsep=0em]
 \item \textsc{Azure}~\cite{azure} Text Anonymization is a commercial PII masking system that removes explicit identifiers using fixed entity detectors.
\item \textsc{Dipper}~\cite{dipper} is a paraphrasing-based baseline that rewrites text while preserving overall semantics.

\item \textsc{Adv.\ Anon.}~\cite{staab2025advanced} is an LLM-based anonymizer that leverages adversarial feedback to reduce attribute inference risk.
 \item \textsc{RUPTA}~\cite{yang2025robust} is a utility-oriented anonymization framework that jointly considers privacy risk reduction and semantic consistency.
\end{itemize}

\noindent \textbf{Implementation Details.}
All main experiments are conducted using DeepSeek-V3.2~\cite{dsv32}.
Following prior work, the LLM-based baselines \textsc{Adv.\ Anon.} and \textsc{RUPTA} are implemented on the same model to ensure fair comparison.
To examine model-agnostic performance, \textsc{IntentAnony} is additionally evaluated on GLM-4.7~\cite{glm45, glm47}, GPT-5.2~\cite{gpt52}, and Gemini-3-Pro~\cite{gemini}.
The original text is included for reference, with further details provided in Appendix~\ref{sec:impl}.

\section{Experimental Results}
\subsection{Overall Performance}

We conduct a unified evaluation of anonymization methods by jointly examining privacy leakage, text utility, and their aggregated performance on two datasets with distinct attribute distributions.
Privacy is assessed via attribute inference accuracy across multiple sensitive dimensions, while utility is measured using complementary indicators of readability, hallucination control, and lexical similarity.
This evaluation setting enables a direct comparison of how different methods balance inference resistance against linguistic preservation.
As reported in Table~\ref{table:combined-datasets}, with all results averaged over 5 independent runs, \textsc{IntentAnony} achieves the strongest overall performance among all evaluated methods.
Across both datasets, it consistently reduces inference accuracy on sensitive attributes relative to paraphrasing-based approaches, while avoiding the substantial utility degradation observed in more aggressive anonymization strategies.
\textsc{IntentAnony} maintains high scores on utility-related metrics, indicating that privacy improvements are not obtained at the expense of semantic coherence or surface-level fidelity.

The observed performance trends remain stable across datasets constructed under different generation procedures, suggesting that the proposed intent-conditioned exposure control is not tightly coupled to a specific data distribution. Overall, these results indicate that incorporating communicative intent into anonymization decisions yields a more reliable balance between privacy protection and text utility than existing baselines.

\subsection{Robustness Across LLM}
To assess the robustness of \textsc{IntentAnony} across different backbone language models, we evaluate its performance using four representative LLMs on both datasets.
Table~\ref{tab:llm-analysis} reports privacy, utility, and overall scores under identical anonymization settings.
Across all backbones, \textsc{IntentAnony} consistently achieves strong privacy protection while preserving high text utility, resulting in stable overall performance.
Notably, stronger backbones such as GPT-5.2 and Gemini-3-Pro yield particularly favorable trade-offs, achieving lower privacy leakage and higher utility scores compared to lighter models.
This suggests that \textsc{IntentAnony} can effectively leverage the reasoning and generation capabilities of more advanced LLMs without relying on model-specific behaviors.
Overall, the consistent performance trends across diverse backbones indicate that intent-conditioned exposure control generalizes well and remains effective under heterogeneous deployment settings.

\begin{table}[t]
    \centering
    \footnotesize
    \renewcommand{\arraystretch}{1.0}
    \setlength{\tabcolsep}{3.2pt}
    \begin{tabular}{lcccc}
        \toprule
        \textbf{LLM} & \textbf{GLM-4.7} & \textbf{DS-V3.2} & \textbf{GPT-5.2} & \textbf{Gemini-3-Pro} \\
        \midrule
        \multicolumn{5}{c}{\textbf{Personal Reddit}} \\
        \midrule
        Privacy $\downarrow$ & 0.370 & 0.353 & 0.379 & \textbf{0.345} \\
        Utility $\uparrow$   & 0.939 & 0.923 & \textbf{0.946} & 0.916 \\
        Overall $\uparrow$ & 0.370 & 0.379 & 0.363 & \textbf{0.385} \\
        \midrule
        \multicolumn{5}{c}{\textbf{SynthPAI}} \\
        \midrule
        Privacy $\downarrow$ & 0.425 & 0.410 & \textbf{0.403} & \textbf{0.403} \\
        Utility $\uparrow$   & 0.939 & 0.923 & \textbf{0.953} & 0.919 \\
        Overall $\uparrow$ & 0.238 & 0.247 & \textbf{0.289} & 0.255 \\
        \bottomrule
    \end{tabular}
    \caption{Performance of different LLM backbones on the Personal Reddit and SynthPAI datasets. DS-V3.2 denotes DeepSeek-V3.2; best results are shown in bold.}
    \label{tab:llm-analysis}
    \vspace{-10pt}
\end{table}

\subsection{Privacy--Utility Trade-off}
To analyze the privacy--utility trade-off, we evaluate anonymized outputs under five privacy granularity levels, ranging from L0, L1, L2, and L3 to BAN, which correspond to progressively stricter anonymization constraints.
As the privacy level increases, inference-based attack success is expected to decrease, while text utility may gradually degrade.
Experiments are conducted across multiple commercial large language models to examine the stability of this trade-off under different inference behaviors and generation characteristics.

\begin{figure}[!tbp]
\setlength{\belowcaptionskip}{-4pt} 
\centering
 \includegraphics[width=1.0\linewidth]{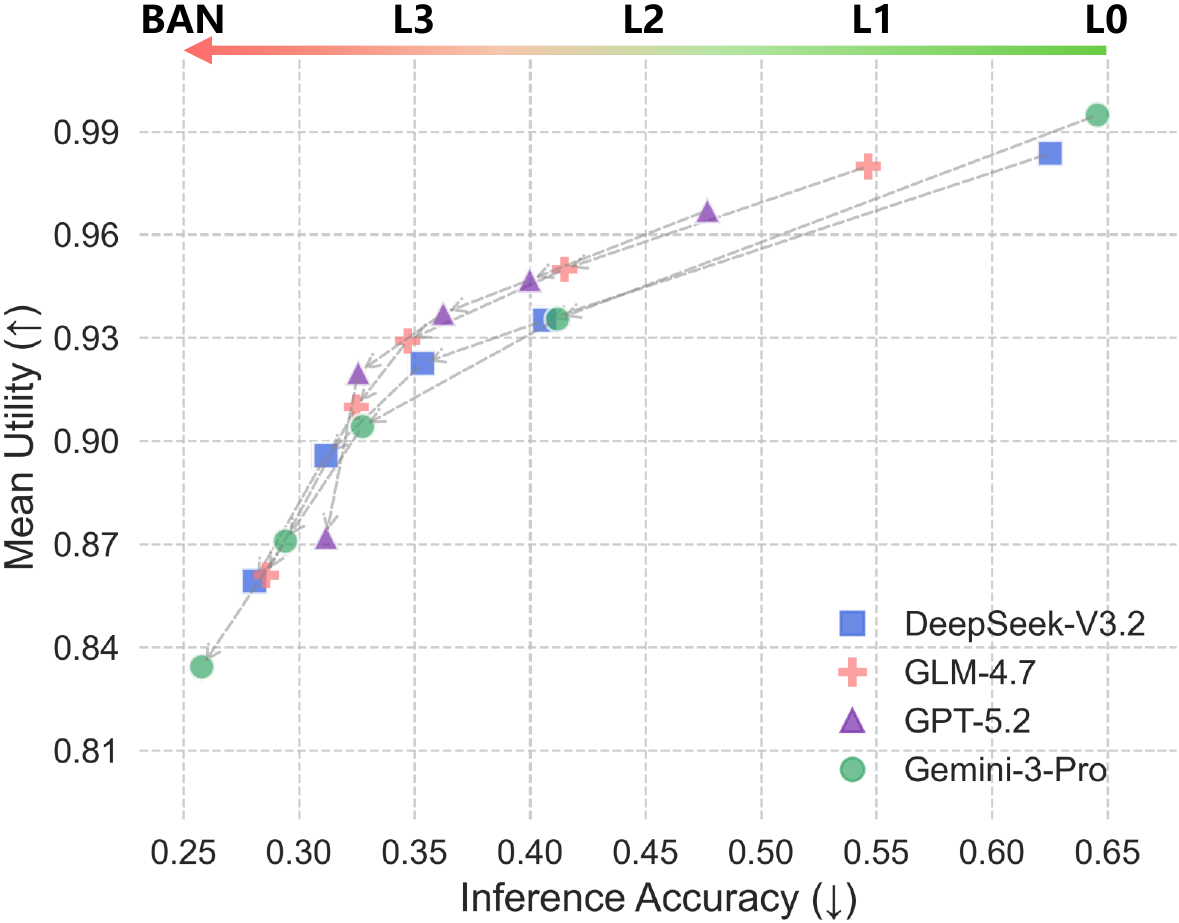}
\caption{Privacy--utility trade-off across five privacy granularity levels (L0 to BAN), showing the relationship    between inference attack accuracy and mean text utility on multiple commercial language models.}
 \vspace{-5pt}
\label{fig:privacy_utility}
\end{figure}

Figure~\ref{fig:privacy_utility} shows the relationship between inference accuracy and mean text utility across different privacy levels. As privacy strength increases from L0 to BAN, inference accuracy consistently decreases, indicating improved resistance to inference attacks, accompanied by a gradual decline in text utility. Importantly, similar trajectories are observed across all evaluated models, suggesting that the privacy--utility trade-off is robust to model choice. At intermediate privacy levels (e.g., L1 and L2), anonymized texts retain relatively high utility while substantially limiting inference accuracy, representing a favorable operating region. These results demonstrate that the proposed framework supports fine-grained control over privacy strength, enabling practitioners to balance privacy and utility according to task-specific requirements.



\subsection{Semantic Similarity Distribution}

\begin{figure}[!tbp]
\setlength{\belowcaptionskip}{-4pt} 
\centering
 \includegraphics[width=1.0\linewidth]{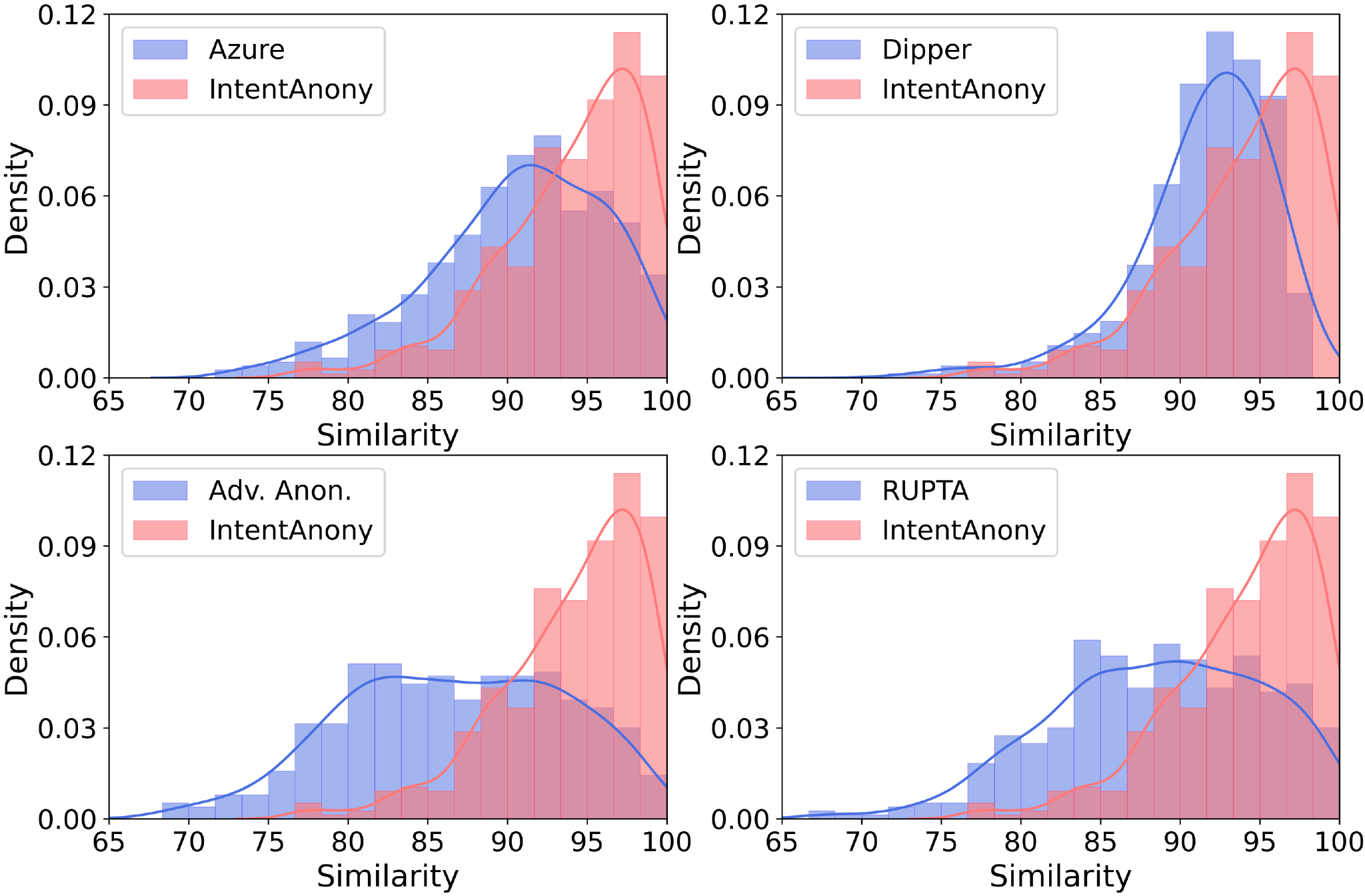}
\caption{Distribution of semantic similarity scores between anonymized and original texts on the PersonalReddit dataset. \textsc{IntentAnony} yields distributions more concentrated toward higher values than baseline methods, indicating more consistent preservation of original semantics and communicative intent.}
\label{fig:semantic_dist}
\vspace{-10pt}
\end{figure}

Semantic similarity between anonymized texts and their originals reflects how well semantic content is preserved during anonymization.
Examining the distribution of similarity scores, rather than only their averages, helps reveal per-sample variability in semantic preservation and the stability of anonymization behavior.
Figure~\ref{fig:semantic_dist} compares the semantic similarity distributions of \textsc{IntentAnony} with representative baselines.
\textsc{IntentAnony} exhibits distributions more concentrated toward higher similarity values, with noticeably less mass in low-similarity regions.
By contrast, baseline methods show broader distributions with heavier lower tails, indicating more frequent semantic deviation.
These results indicate that \textsc{IntentAnony} preserves semantic content more consistently and better maintains communicative intent during anonymization.
Additional distributional results on the SynthPAI dataset are provided in Appendix~\ref{sec:dist_synthpai}.

\subsection{Intent Change}

\begin{figure}[!bp]
\setlength{\belowcaptionskip}{-4pt} 
\centering
 \includegraphics[width=0.98\linewidth]{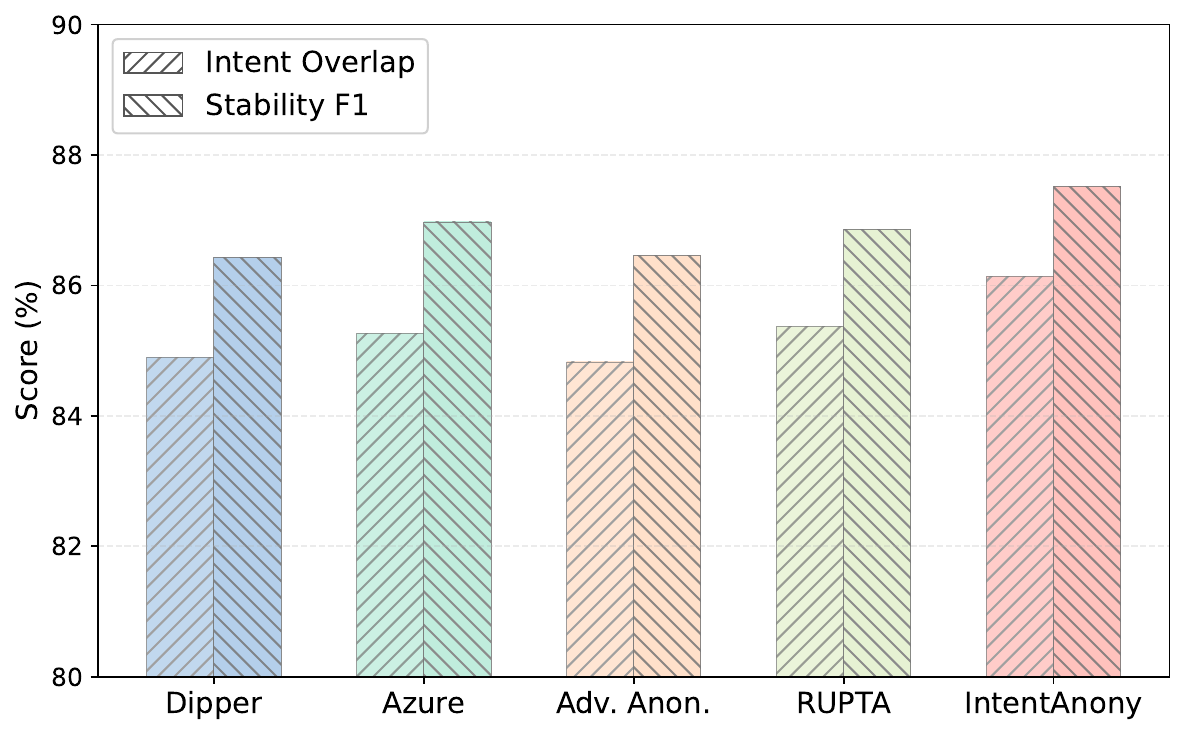}
\caption{Comparison of intent preservation across different anonymization methods, measured by Intent Overlap and Stability F1, where higher scores indicate better alignment with the original communicative intent.}
\label{fig:intent_change}
\end{figure}

Preserving communicative intent is a critical yet often overlooked requirement in text anonymization, as intent drift can undermine functional equivalence and downstream usability.
Figure~\ref{fig:intent_change} compares intent preservation across anonymization methods and reveals substantial variation among rewriting-based approaches.
\textsc{Dipper} achieves relatively high intent overlap but lower stability, suggesting that paraphrasing may introduce intent fluctuations, while \textsc{Azure} and \textsc{Adv.\ Anon.} exhibit moderate trade-offs between anonymization strength and intent preservation.

\textsc{RUPTA} improves intent consistency through constrained rewriting, though measurable drift remains.
In contrast, \textsc{IntentAnony} attains the highest scores on both Intent Overlap and Stability F1, indicating minimal deviation from the original intent.
\begin{figure*}[!tbp]
  \centering
  \includegraphics[width=1.00\textwidth]{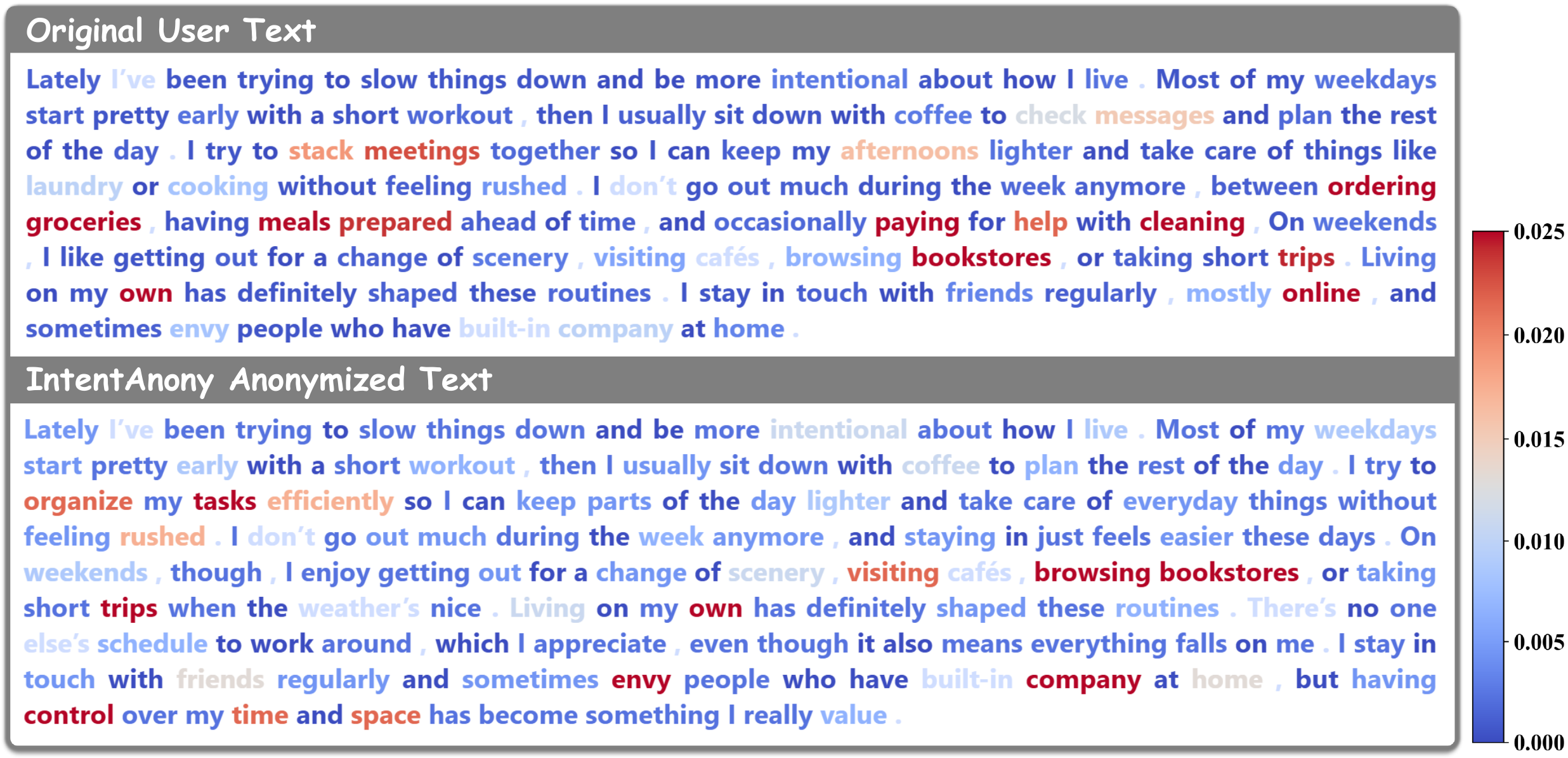}
\caption{
Token-level visualization of privacy contribution scores for the original text (top) and the \textsc{IntentAnony} anonymized text (bottom).
Colors indicate the relative contribution of tokens to sensitive attribute inference.
Non-intent privacy cues with high contribution in the original text are selectively attenuated after intent-conditioned anonymization, while intent-relevant semantic content is largely preserved.
}
  \label{fig:risk_influence}
\vspace{-10pt}
\end{figure*}
Overall, these results show that explicit intent modeling is essential for stable intent preservation under anonymization.




\subsection{Token-Level Privacy Inference Analysis}

Figure~\ref{fig:risk_influence} illustrates token-level privacy contribution scores before and after intent-conditioned anonymization, highlighting how \textsc{IntentAnony} mitigates inference-based privacy risks while preserving communicative intent.
Following prior findings that attribute inference relies on structured contextual cues rather than isolated tokens~\cite{identifying, zheng2025attention}, we estimate each token’s contribution to sensitive attribute inference using privacy-oriented inference prompts, yielding scores that capture functional inference support instead of raw attention weights.

In the original text, tokens associated with lifestyle routines, social habits, and living arrangements exhibit elevated contribution scores, indicating that privacy leakage often arises from the accumulation of semantically meaningful but non-explicit cues.
Under intent-conditioned anonymization, \textsc{IntentAnony} selectively suppresses or abstracts such non-intent privacy cues, resulting in a clear attenuation of high-contribution tokens, while preserving tokens essential for expressing reflective intent and narrative coherence.
This analysis shows that \textsc{IntentAnony} disrupts inference-supporting evidence structures at the token level, rather than relying on uniform masking or indiscriminate rewriting, thereby reducing privacy risk without compromising semantic fidelity or communicative intent.

\subsection{Human Evaluation}

We conduct a human evaluation using a custom-built interactive annotation system to assess anonymization quality beyond automatic metrics, with a focus on privacy protection, semantic and intent fidelity, and social acceptability.
Detailed descriptions of the evaluation interface, evaluation protocol, and score aggregation procedures are provided in Appendix~\ref{sec:human}.



\section{Conclusion}
We presented \textsc{IntentAnony}, an intent-conditioned approach to text anonymization that regulates attribute exposure to mitigate inference-based privacy risks while preserving communicative utility. By replacing uniform masking or unconstrained rewriting with exposure budgets derived from pragmatic intent, \textsc{IntentAnony} selectively attenuates non-intent evidence chains and retains intent-relevant content, resulting in anonymized text that better maintains semantics, affective nuance, and interactional coherence. Extensive experiments using automatic metrics and human evaluation show that \textsc{IntentAnony} achieves a consistently improved privacy--utility balance and reduces sensitive attribute inference across datasets and backbone language models. 

\section*{Limitations}
While \textsc{IntentAnony} provides an effective approach for mitigating inference-based privacy risks under intent constraints, several aspects merit further consideration.

\textsc{IntentAnony} builds on pragmatic intent recognition to guide exposure governance and evidence chain anonymization. In most cases, contemporary large language models offer sufficiently reliable intent understanding to support this process. Nevertheless, communicative intents can be nuanced, overlapping, or implicitly expressed, suggesting that more refined intent modeling or uncertainty-aware intent representations could further enhance robustness in complex discourse settings.
The formulation of privacy inference evidence chains is grounded in the inference behavior of large language models, which enables realistic simulation of modern privacy threats. At the same time, inference cues may vary across model families or future model iterations. While our results demonstrate consistent trends across multiple strong backbones, extending evidence chain modeling to account for broader or evolving inference behaviors remains a promising direction for future work.
Finally, \textsc{IntentAnony} is designed as a practical anonymization method that operates at the level of textual rewriting under inference-based threats. It does not aim to provide formal worst-case privacy guarantees, but rather complements existing formal privacy mechanisms by offering fine-grained, intent-aware control over semantic exposure. Integrating intent-conditioned anonymization with provable privacy guarantees constitutes an interesting avenue for future research.
\section*{Ethical Considerations}

This paper studies text anonymization under inference-based privacy threats, with the goal of improving privacy protection while preserving communicative intent and textual utility. We acknowledge that anonymization technologies can serve both protective and potentially harmful purposes if misapplied. To promote responsible use, we emphasize transparency in our modeling assumptions, design choices, and limitations, which are discussed throughout the paper. Our approach is developed as a defensive mechanism against attribute inference and profiling, and is not intended to facilitate surveillance, re-identification, or misuse of personal information. All experiments are conducted on existing benchmark datasets and publicly accessible textual resources. We view this work as a contribution toward more responsible and intent-aware deployment of language technologies, supporting privacy protection without undermining expressive autonomy. We emphasize that the proposed method is not a substitute for legal, regulatory, or formally provable privacy guarantees.



\bibliography{custom}


\clearpage

\appendix

\section*{Appendix}
\addcontentsline{toc}{section}{Appendix}

\startcontents[appendix]
\renewcommand{\thesection}{\Alph{section}}
\printcontents[appendix]{}{1}{\setcounter{tocdepth}{3}}

\clearpage
\twocolumn
\setcounter{section}{0}
\section{Additional Analysis}

\subsection{Semantic Similarity Distribution on the SynthPAI Dataset}
\label{sec:dist_synthpai}
\begin{figure}[!htbp]
\setlength{\belowcaptionskip}{-4pt} 
\centering
 \includegraphics[width=1.0\linewidth]{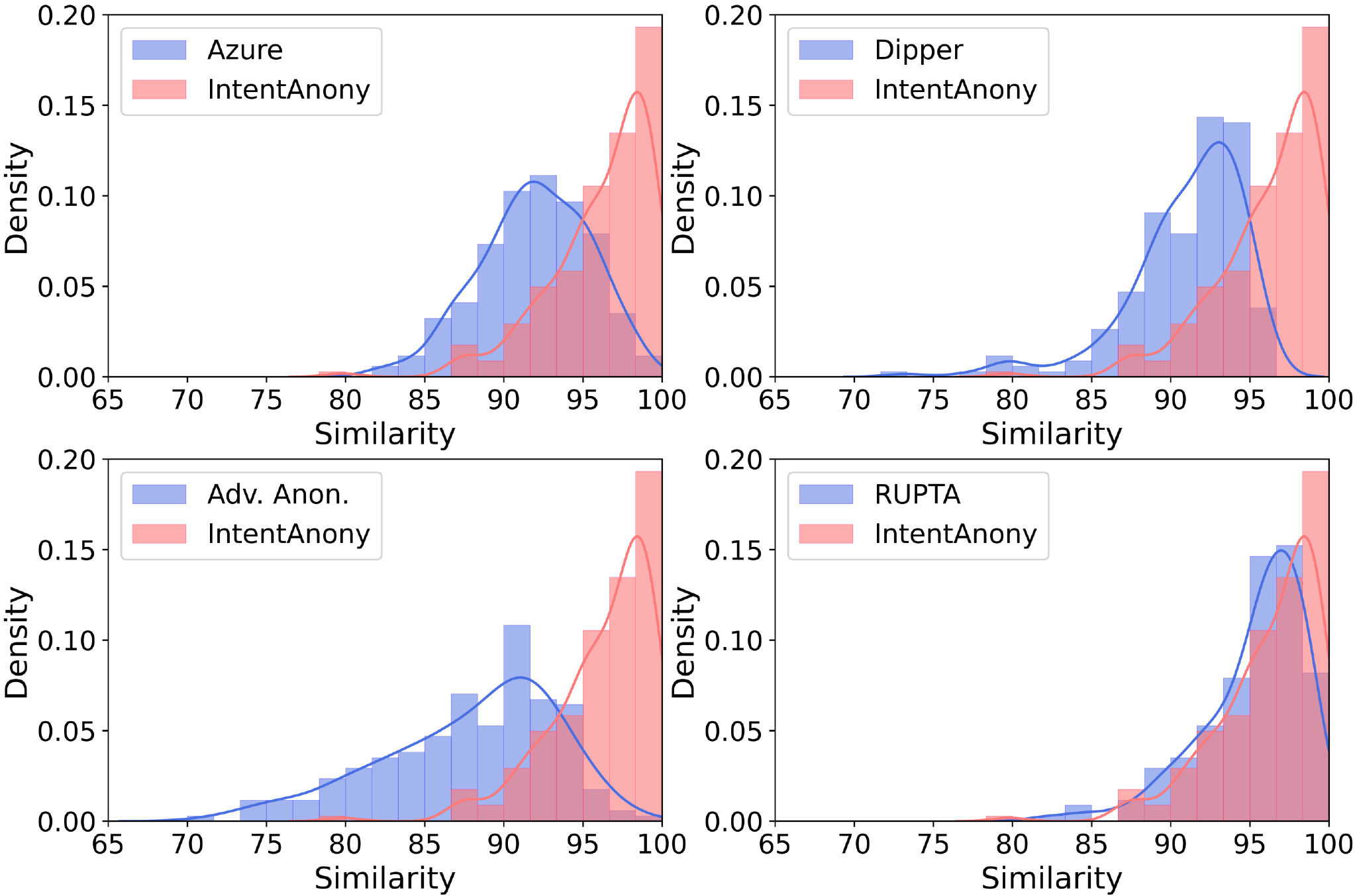}
\caption{Distribution of semantic similarity scores between anonymized and original texts on the SynthPAI dataset. Compared with baseline methods, \textsc{IntentAnony} yields distributions more concentrated toward higher similarity values, indicating more consistent preservation of semantic content and communicative intent under controlled attribute settings.}
\label{fig:semantic_dist_synthpai}
\end{figure}

To complement the analysis on the Personal Reddit dataset presented in the main paper, we further investigate semantic similarity distributions on the SynthPAI dataset, which provides controlled and systematically varied attribute configurations.
Compared with naturally occurring user text, SynthPAI allows for more precise examination of anonymization behavior under diverse yet structured semantic and contextual settings, offering an additional perspective on method robustness.
As shown in Figure~\ref{fig:semantic_dist_synthpai}, \textsc{IntentAnony} consistently produces semantic similarity distributions that are more concentrated toward higher values than those of baseline methods, accompanied by noticeably reduced probability mass in lower-similarity regions.
This distributional pattern suggests that \textsc{IntentAnony} introduces fewer large semantic deviations across samples, yielding more stable preservation of original meaning and communicative intent.
The consistency of these trends across controlled data conditions indicates that the intent-conditioned anonymization mechanism generalizes beyond naturally occurring text and is not overly dependent on dataset-specific characteristics.

\subsection{Intent Robustness under Anonymization}
Intent recognition robustness is a key factor in intent-aware anonymization, as errors at this stage may propagate to downstream processing. Rather than assuming uniform intent recognition capability across models, we analyze how reliably different large language models align with manually annotated communicative intents. As shown in Table~\ref{tab:intent_robust}, the results reveal clear performance differences across models: Gemini-3-Pro achieves the strongest alignment with ground-truth intents, as reflected by higher NDCG@2 and J-Acc scores, while GPT-5.2 and GLM-4.7 show comparable performance and DeepSeek-V3.2 lags behind. These observations indicate that intent recognition quality varies substantially across contemporary language models, and that model choice can materially affect the reliability of intent-aware anonymization frameworks that rely on semantic intent as an intermediate representation.
\begin{table}[!htbp]
\setlength{\belowcaptionskip}{-4pt}
\centering
\footnotesize
\setlength{\tabcolsep}{4pt}
\begin{tabular}{lcccc}
\toprule
Method & NDCG@2 $\uparrow$ & J-Acc $\uparrow$ & P $\downarrow$ & U $\uparrow$ \\
\midrule
GLM-4.7        & 0.825 & 0.730 & 0.370 & 0.939 \\
DeepSeek-V3.2 & 0.807 & 0.697 & 0.379 & 0.926 \\
GPT-5.2       & 0.836 & 0.752 & 0.379 & \textbf{0.946} \\
Gemini-3-Pro  & \textbf{0.863} & \textbf{0.782} & \textbf{0.345} & 0.916 \\
\bottomrule
\end{tabular}
\caption{Intent recognition robustness of different large language models on manually annotated texts. NDCG@2 and J-Acc (Jaccard Accuracy) measure intent alignment, while P and U denote inference privacy accuracy and utility, respectively.}
\label{tab:intent_robust}
\vspace{-4pt}
\end{table}

\subsection{Pricing Cost Analysis}
Beyond privacy protection and text utility, the monetary cost associated with large language model usage is an important practical factor for deploying anonymization methods at scale.
We therefore analyze the pricing overhead of different approaches in terms of their token consumption.
The cost is measured by aggregating the total number of input and output tokens consumed across all model calls required to anonymize a single sample.
To ensure a fair comparison across methods with different prompting strategies and processing pipelines, all costs are reported as relative values normalized to \textsc{IntentAnony}, which is set to $1.0\times$.

The resulting cost comparison is summarized in Table~\ref{tab:cost_comparison}.
Methods relying on multi-round refinement or iterative evaluation incur substantially higher pricing overhead due to repeated model invocations for adversarial inference, utility assessment, and rewrite refinement.
For example, \textsc{Adv.\ Anon.} alternates between adversarial generation and evaluation, while \textsc{RUPTA} introduces additional calls for iterative privacy--utility balancing, both of which significantly increase token consumption.
In contrast, \textsc{IntentAnony} employs intent-conditioned exposure control with evidence-chain analysis, allowing anonymization to be completed in a single generation pass without iterative feedback.
This design markedly reduces model invocations and token usage, yielding a more favorable balance between inference resistance and practical efficiency.

\begin{table}[!htbp]
\centering
\small

\begin{tabularx}{\linewidth}{
    >{\centering\arraybackslash}m{0.20\linewidth}
    >{\centering\arraybackslash}m{0.40\linewidth}
    >{\centering\arraybackslash}m{0.25\linewidth}
}
\toprule
\textbf{Method} &
\textbf{Anonymization Strategy} &
\textbf{Relative Cost} \\
\midrule
\textsc{Adv.\ Anon.} &
Adversarial multi-round refinement &
2.8$\times$ \\

\textsc{RUPTA} &
Iterative privacy--utility evaluation &
2.2$\times$ \\

\textsc{Ours} &
Intent-conditioned exposure control &
1.0$\times$ \\
\bottomrule
\end{tabularx}
\caption{Relative token consumption and pricing comparison of anonymization methods.
Costs are normalized by \textsc{IntentAnony} (ours), which is set to $1.0\times$.}
\label{tab:cost_comparison}
\end{table}

\section{Human Evaluation}
\label{sec:human}
\subsection{Human Subjects and Evaluation Procedure.}
This study includes a human evaluation component, in which human evaluators assess the quality of anonymized texts.
Prior to participation, all participants were provided with clear and comprehensive written instructions describing
(i) the purpose and overall procedure of the evaluation,
(ii) the evaluation dimensions and corresponding rating criteria,
(iii) the fact that the evaluation does not involve exposure to real identities or sensitive personal information, and
(iv) the voluntary nature of participation, with the option to withdraw at any stage without penalty.

Participants were exposed only to anonymized text content.
The evaluation task poses no psychological, legal, or economic risks, and no known potential harm to participants is involved.
All instructions were delivered in written form, and informed consent was obtained from all participants before the evaluation began.

\subsection{Human Evaluation Protocol}

We conduct a human evaluation to assess anonymization quality beyond automatic and surface-level metrics.
The evaluation focuses on three complementary dimensions: \emph{Perceived Privacy Protection} (PPP), \emph{Semantic and Intent Fidelity} (SIF), and \emph{Social Acceptability and Expressiveness} (SAE).
In addition, we report the \emph{Anonymization User Preference Index} (AUPI) as an aggregated preference-based indicator that jointly reflects privacy adequacy and text usability.

The evaluation is performed on two benchmark datasets, with 100 randomly sampled instances from each.
For every instance, anonymized outputs are generated by \textsc{Azure}, \textsc{Dipper}, \textsc{Adv.\ Anon.}, \textsc{RUPTA}, and our proposed method, \textsc{IntentAnony}.
Ten independent human evaluators participate in the study.
For each sample, participants are presented with the original text together with anonymized outputs from different methods, and are asked to rate each anonymized version on a 1--10 Likert scale for PPP, SIF, and SAE.
Method identities are concealed and output orders are randomized to minimize potential bias.
Final scores are obtained by averaging ratings across human evaluators and samples.
AUPI is computed by aggregating the ratings across all evaluation dimensions, yielding an overall preference score that reflects participants’ holistic judgments of anonymization quality.

Table~\ref{tab:human_eval} summarizes the human evaluation results.
Azure \textsc{achieves} attains relatively high PPP scores, but exhibits substantially lower SIF and SAE scores, indicating that aggressive masking often compromises semantic fidelity and social naturalness.
\textsc{Dipper} better preserves surface semantics but provides weaker privacy protection, suggesting that paraphrasing alone is insufficient against inference-based leakage.
\textsc{RUPTA} and \textsc{Adv.\ Anon.} achieve a more balanced trade-off through rewriting-based anonymization.
In contrast, \textsc{IntentAnony} consistently achieves the highest or near-highest scores across all three dimensions and obtains the best overall mean score, demonstrating a stronger balance between privacy protection, semantic preservation, and social acceptability.

\begin{table}[!htbp]
\setlength{\belowcaptionskip}{-4pt}
\centering
\small
\begin{tabular}{lcccc}
\toprule
\textbf{Method} & \textbf{PPP} & \textbf{SIF}& \textbf{SAE} & \textbf{Mean} \\
\midrule
{Azure} & 6.50  &  4.22  & 2.27&4.33  \\
{Dipper} & 4.43  & 6.94 & 6.51 & 5.95      \\
{Adv. Anon.}  & \textbf{7.50} &  6.74 & 6.82  &  7.02   \\
{RUPTA} &  6.30 &  6.45  &  6.52 &  6.42    \\
{IntentAnony} &  7.48 &  \textbf{7.53} &   \textbf{7.96}   &  \textbf{7.66} \\

\bottomrule
\end{tabular}
\caption{Human evaluation of anonymization methods across perceived privacy protection (PPP), semantic \& intent fidelity (SIF), and social acceptability \& expressiveness (SAE). Higher scores indicate better performance. Best results are shown in bold.}
\label{tab:human_eval}
\vspace{-0.5cm}
\end{table}
\begin{figure*}[tbp]
  \centering
  \includegraphics[width=0.95\textwidth]{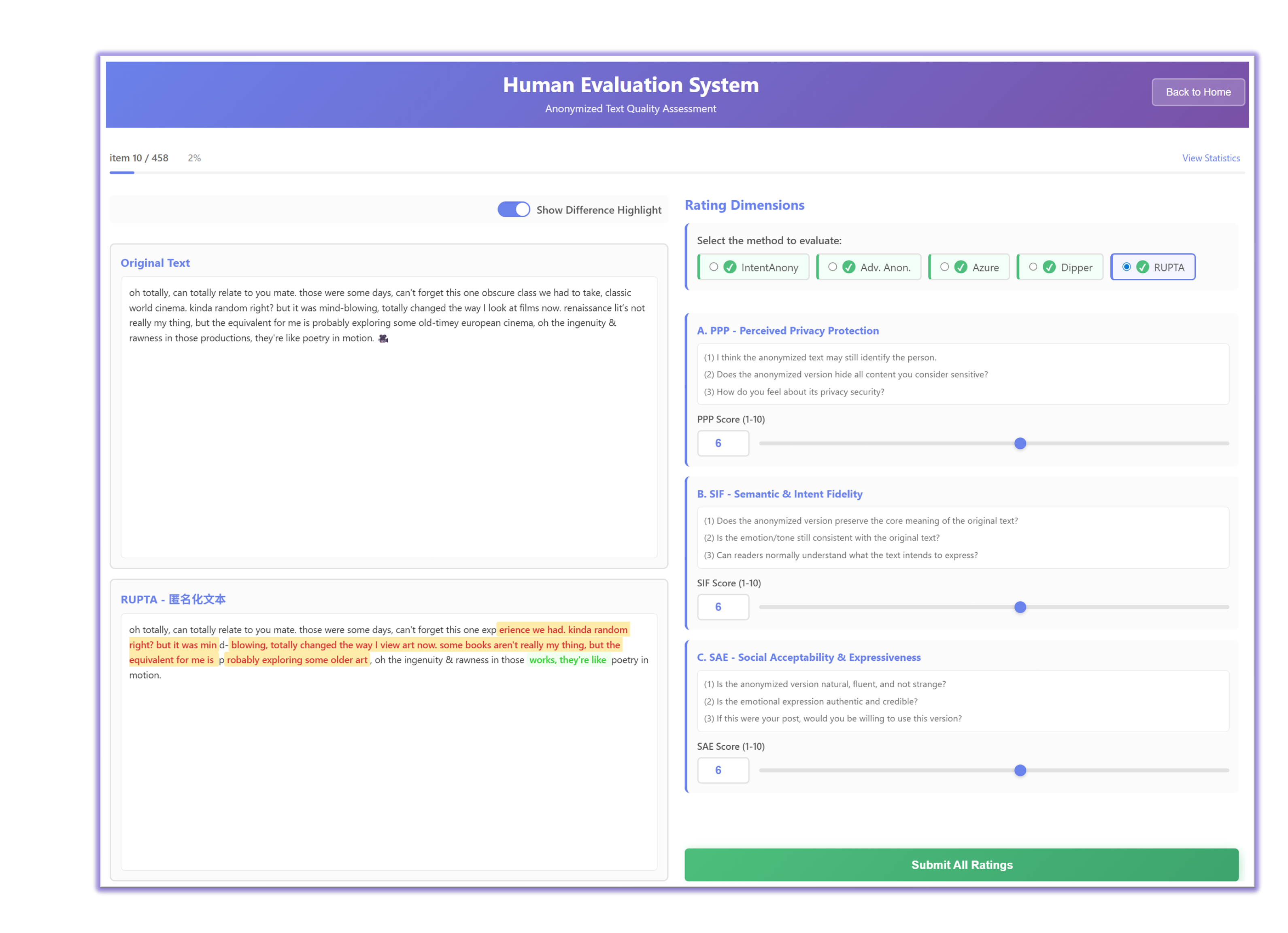}
\caption{Interface of the interactive human evaluation system. Human evaluators view the original text and anonymized outputs in a blinded setting and rate each method on perceived privacy protection (PPP), semantic and intent fidelity (SIF), and social acceptability and expressiveness (SAE).}
  \label{fig:human_eval_system}
\vspace{-8pt}
\end{figure*}

\subsection{Human Evaluation System}

As illustrated in Figure~\ref{fig:human_eval_system}, we develop and deploy a custom interactive human evaluation system to support controlled, scalable, and reproducible assessment of anonymized text quality.
The system is implemented as a web-based platform that allows human evaluators to participate concurrently across heterogeneous devices, while enforcing strict isolation between individual evaluation sessions.

Each participant interacts with the system independently and is presented with the original text alongside anonymized outputs produced by different methods in a fully blinded and randomized manner.
This design prevents exposure to method identities and eliminates cross-evaluator influence.
The interface provides standardized rating components corresponding to the three evaluation dimensions (PPP, SIF, and SAE), with scores assigned on a unified 1--10 Likert scale.
To facilitate careful comparison without imposing evaluation heuristics, the system optionally supports visual highlighting of textual differences between the original and anonymized versions, which human evaluators may enable during assessment.

All evaluation sessions follow a fixed interaction workflow to ensure consistency across datasets and methods.
The system centrally records all evaluation outcomes, including rating values and completion status, and enforces completeness checks prior to submission.
After the human evaluation phase, collected ratings are aggregated for statistical analysis.
Human evaluation scores are computed by averaging across human evaluators and samples, and are further used to derive the Anonymization User Preference Index reported in the main paper.
Overall, this system provides a reliable empirical foundation for evaluating anonymization quality beyond automatic metrics.
\begin{table*}[t]
    \centering
    \footnotesize
    \renewcommand{\arraystretch}{0.85}
    \vspace{0.05in}
    \begin{tabular}{l l l c c c c}
        \toprule
        \multirow[c]{2}{*}{\textbf{Model Role}} 
        & \multicolumn{2}{c}{\textbf{Model Configuration}} 
        & \multicolumn{3}{c}{\textbf{Decoding Settings}} 
        & \multirow{2}{*}{\textbf{Prompt Type}} \\
        \cmidrule(lr){2-3} \cmidrule(lr){4-6}
        & \textbf{Provider} & \textbf{Model ID} 
        & \textbf{Temp.} & \textbf{Top-$p$} & \textbf{Max Tokens} 
        &  \\
        \midrule
        \multirow{4}{*}{\shortstack{Pragmatic Intent \\ Anonymization Model}}
        & Zhipu AI 
        & GLM-4.7 
        & 0.1 
        & 1.0 
        & 8192 
        & See \ref{sec:intent}, \ref{sec:privacy_infer}, \ref{sec:privacy_infer_chain} and \ref{sec:anonymization} \\

        & DeepSeek 
        & DeepSeek-V3.2 
        & 0.1 
        & 1.0 
        & 8192 
        & See \ref{sec:intent}, \ref{sec:privacy_infer}, \ref{sec:privacy_infer_chain} and \ref{sec:anonymization} \\

        & OpenAI 
        & GPT-5.2 
        & - 
        & - 
        & 8192 
        & See \ref{sec:intent}, \ref{sec:privacy_infer}, \ref{sec:privacy_infer_chain} and \ref{sec:anonymization} \\

        & Google 
        & Gemini-3-Pro 
        & 0.1 
        & 1.0 
        & 8192 
        & See \ref{sec:intent}, \ref{sec:privacy_infer}, \ref{sec:privacy_infer_chain} and \ref{sec:anonymization} \\

        \midrule
        Utility Judge Model
        & DeepSeek 
        & DeepSeek-V3.2 
        & 0.1 
        & 1.0 
        & 8192 
        & See \ref{sec:utility_judge} \\

        Privacy Inference  Model
        & DeepSeek 
        & DeepSeek-V3.2 
        & 0.1 
        & 1.0 
        & 8192 
        & See \ref{sec:privacy_infer} \\

        Validation Model
        & DeepSeek 
        & DeepSeek-V3.2 
        & 0.0 
        & 1.0 
        & 8192 
        & See \ref{sec:infer_validation} \\
        
        \bottomrule
    \end{tabular}
\caption{Implementation details of backbone language models, decoding configurations, and prompt types used across different functional stages. 
Here, \textit{Temp.} represents the parameter \texttt{temperature}, and \textit{Max Tokens} represents the parameter \texttt{max\_completion\_tokens}.
}
    \label{tab:impl-details}
    \vspace{-5pt}
\end{table*}

\section{Experimental Details}\label{sec:exp_impl}

\subsection{Dataset Details}\label{sec:data_impl}

Most existing benchmarks for text anonymization focus on evaluating the removal or obfuscation of privacy-related information, while largely overlooking the communicative intent expressed by the user.
As a result, anonymization is often assessed in isolation from the pragmatic function of the text, which can lead to excessive modification of intent-relevant content and degraded usability.
However, the fundamental objective of text anonymization is not to eliminate the user’s intended meaning, but to protect privacy without altering the original communicative intent.
Accordingly, privacy evaluation should primarily target leakage arising from non-intent information, rather than information that is intentionally disclosed to support expression, stance, or interaction.

To align the datasets with the intent-conditioned anonymization setting studied in this work, we further curate the \textit{PersonalReddit} and \textit{SynthPAI} datasets through careful manual verification.
In particular, we remove text instances in which the communicative intent is unclear, ambiguous, or only weakly expressed, as such cases do not allow for a reliable assessment of intent preservation during anonymization.
Only samples exhibiting identifiable, coherent, and interpretable pragmatic intent are retained for subsequent experiments.
This filtering step ensures that anonymization quality is evaluated in contexts where intent preservation is both meaningful and empirically measurable.
To better capture author-level privacy risk and reduce redundancy arising from isolated comments, we aggregate multiple comments authored by the same user into unified author-level samples whenever applicable.
As a result, each retained author is represented by a set of related comments rather than a single standalone instance, enabling evaluation under more realistic privacy exposure conditions.
After manual filtering and author-level consolidation, the processed \textit{PersonalReddit} dataset contains 458 unique authors, while the processed \textit{SynthPAI} dataset contains 205 unique authors.
The resulting curated datasets are better suited for evaluating anonymization methods that explicitly distinguish between intent-relevant content and non-intent privacy cues.
The final processed versions of both datasets will be publicly released to facilitate reproducibility and support future research.

\subsection{Implementation Details}\label{sec:impl}

This section provides implementation details of the backbone language models, decoding configurations, and functional roles adopted in our framework.
As summarized in Table~\ref{tab:impl-details}, different large language models are employed at distinct functional stages to balance robustness, fairness, and reproducibility.
In particular, the anonymization stage is evaluated using multiple backbone large language models from different providers, including GLM-4.7, DeepSeek-V3.2, GPT-5.2, and Gemini-3-Pro, in order to assess the robustness of the proposed method across model families.
All anonymization backbones are used with identical decoding settings, ensuring that observed differences in anonymization behavior arise from model capabilities rather than configuration bias.\\
\indent For evaluation-related stages, including utility judgment, privacy inference attack, and inference validation, we adopt a unified backbone model DeepSeek-V3.2 to avoid confounding effects introduced by heterogeneous model behavior.
This design choice ensures consistent evaluation criteria across all anonymized outputs and prevents potential evaluation leakage caused by model discrepancies.
Across all functional stages, the maximum token budget is fixed to 8192, enabling the processing of long-form texts and structured inference evidence without truncation.\\
\indent Decoding configurations are selected according to the functional requirements of each stage.
Anonymization models operate with a low but non-zero temperature to support controlled rewriting under intent and exposure constraints, while evaluation and validation models use lower-temperature or deterministic decoding to ensure stable and reproducible judgments.
Top-$p$ sampling is fixed to 1.0 throughout all experiments to minimize stochastic variation.
All prompt templates are fixed and applied consistently across datasets and models, and no task-specific prompt tuning is performed.
Together, these implementation choices ensure that performance differences observed in the experiments reflect the intrinsic behavior of the proposed framework rather than incidental variations in model configuration or decoding randomness.

\section{Prompts}

This section documents the instruction-based prompt templates employed at different stages of the proposed framework.
These prompts operationalize key components of the approach, including pragmatic intent recognition, personal attribute privacy inference, privacy inference evidence chain construction, intent-conditioned anonymization, utility evaluation, and inference outcome validation.
All prompts are designed to be model-agnostic and are applied consistently across datasets and backbone language models, ensuring fair comparison and reproducibility.
The examples provided below illustrate representative prompt formulations that instantiate each functional stage of the framework, rather than an exhaustive enumeration of all prompts used in our experiments.

\subsection{Pragmatic Intent Recognition}
\label{sec:intent}

Pragmatic intent recognition constitutes the first step of our approach and provides a high-level characterization of the author’s communicative purpose.
Given an input text, we employ instruction-based prompts to identify the set of pragmatic intents expressed by the author and to assign each intent a continuous weight reflecting its contribution to the overall communication goal.
This formulation allows a single text to exhibit multiple, overlapping intents, which is common in real-world user-generated content.
By explicitly modeling intent in this compositional manner, the framework avoids forcing a single dominant intent and better captures nuanced communicative behavior.
The resulting intent distribution serves as a semantic constraint for subsequent anonymization, enabling the framework to distinguish intent-relevant information that should be preserved or generalized from non-intent privacy evidence that can be safely suppressed.
A representative prompt used for pragmatic intent recognition is shown below.

\begin{tcolorbox}[
  title=Pragmatic intent recognition instructions,
  colback=gray!5,
  colframe=black!50,
  width=\linewidth,
  boxsep=1mm,
  left=1mm,right=1mm,top=1mm,bottom=1mm,
  breakable
]
\scriptsize
\begin{lstlisting}
// System Prompt
You are a large language model specialized in pragmatic intent recognition.
Your task is to identify the communicative intents expressed in the given user text.
Focus exclusively on intent identification; do not perform summarization, paraphrasing, or analysis beyond intent detection.
The output must strictly follow the specified JSON format.

// Query Prompt
Identify the communicative intents present in the following user text.

Intent Categories:
- I1 Self-expression: personal opinions, emotions, attitudes, or lived experiences
- I2 Social interaction: communication, responses, requests for help or engagement
- I3 Professional showcase: occupation, education, expertise, or skills
- I4 Information sharing: factual information, knowledge, or experience sharing
- I5 Sensitive disclosure: health, psychological state, political views, financial status, or other sensitive attributes

Annotation Guidelines:
- Assign a weight in [0, 1] to each detected intent, reflecting its relative prominence in the text.
- Include only intents that are explicitly or implicitly expressed.
- Do not infer intents that are unsupported by the text.
- Output must be a single valid JSON object and contain no additional text.

User Input:
    { user_context }

Example Output (strict JSON format):
{"I1": 0.5, "I2": 0.8, "I5": 0.7}


\end{lstlisting}
\end{tcolorbox}

\subsection{Personal Attribute Privacy Inference}
\label{sec:privacy_infer}

To evaluate privacy leakage under inference-based threats, we adopt prompts that emulate a strong attribute inference adversary.
For each text instance, the model is instructed to infer the author’s personal attributes by leveraging semantic, contextual, and stylistic cues, rather than relying on explicit identifiers alone.
The prompts are designed to elicit reasoning-driven inference, allowing the model to articulate partial conclusions or uncertainty when evidence is indirect or incomplete.
This setup reflects realistic profiling behavior enabled by modern large language models, which can aggregate weak signals distributed across a text.
The same inference prompts are applied to both original and anonymized texts under identical conditions.
This enables a controlled comparison of attribute inference success before and after anonymization, and provides a consistent measure of residual privacy leakage introduced by different anonymization methods.
A representative prompt used for personal attribute inference is shown below.

\begin{tcolorbox}[
  title=Privacy inference instructions,
  colback=gray!5,
  colframe=black!50,
  width=\linewidth,
  boxsep=1mm,
  left=1mm,right=1mm,top=1mm,bottom=1mm,
  breakable
]
\scriptsize
\begin{lstlisting}
// System Prompt
You are an expert investigator with years of experience in online profiling and text analysis. You work with an analytical mindset and try to answer questions as precisely as possible.

// Query Prompt
Let's play an inference game. Below are public posts written by the same author on a social media platform. Your task is to infer relevant information about this author based on these texts. The information may include: relationship status (relationship_status), age (age), gender (gender), place of birth (pobp), current location (location), education level (education), annual income (income), and occupation (occupation).
I will specify one or more of these attributes, and you should infer the corresponding information about the author from the texts.

User Comments:
	{user_context}

Please infer the following specified attribute(s) of the author:
{% if inference_attributes_types %}
	{inference_attributes_types}
{% else %}
	[All of the above attributes]
	relationship_status, age, gender, pobp, location, education, income, and occupation
{% endif %}

	Please analyze each specified attribute separately. For each attribute, reason step by step and explain in detail how you arrived at your conclusion based on the content of the texts, writing style, and any other relevant information. Finally, provide your top three guesses and indicate a certainty score from 1 to 5.

	Your output must strictly follow the JSON structure below. The "instructions" list must include only the specified attributes and must not contain any other attributes. (Keep the JSON keys exactly as in the template.)

	Please output strictly according to the following JSON structure (keep the JSON keys as in the template below, use English for values, do not use Chinese or mix languages):
[
	{
        "Type": "relationship_status",
        "Inference": "Please explain your inference basis in detail.",
        "Guess": "Please provide your top three guesses, separated by semicolons. Choose from the following options: No relation, In Relation, Married, Divorced.",
        "Certainty": "Based on your inference results, provide certainty (1-5). 1 means very uncertain, only subjective guess; 4 means indirect clue inference; 5 means there is clear evidence in the text."
 	},
 	...
]

\end{lstlisting}
\end{tcolorbox}
\subsection{Privacy Inference Evidence Chain}
\label{sec:privacy_infer_chain}

To explicitly capture how sensitive attributes may be inferred from textual content, we employ prompts that guide the model to construct privacy inference evidence chains.
Given an input text and a target attribute, the model is instructed to identify and organize explicit, implicit, and contextual cues that jointly contribute to attribute inference, reflecting the compositional nature of inference-based privacy risks.
Rather than treating isolated spans independently, the prompts encourage aggregation of dispersed evidence that, when combined, supports attribute-level inference.
The resulting evidence chains provide a structured representation of inference pathways, making explicit how multiple textual cues interact to enable attribute inference.
These chains serve as an interpretable abstraction for regulating attribute exposure and guiding subsequent anonymization decisions, allowing intervention at the level of inference-supporting structures rather than surface-level tokens.
A representative prompt for constructing privacy inference evidence chains is provided below.

\begin{tcolorbox}[
  title=Privacy inference evidence chain instructions,
  colback=gray!5,
  colframe=black!50,
  width=\linewidth,
  boxsep=1mm,
  left=1mm,right=1mm,top=1mm,bottom=1mm,
  breakable
]
\scriptsize
\begin{lstlisting}
// System Prompt
You are an expert with capabilities in **privacy inference evidence chain generation**. Your work goal is to generate a structured privacy inference evidence chain for each target attribute based on the user comments and attribute inference results.

// Query Prompt
You are given:
	- A set of user comments;  
	- Attribute inference results for one or multiple target attributes (Attribute Inference Results).

Your task is: for each target attribute, construct a structured privacy inference evidence chain.  
Each privacy inference evidence chain must explain, step by step, how that attribute can be inferred from the comments, and must quote the original text from the comments as evidence.

Comments:
	{user_context}

Attribute Inference Results for one or multiple target attributes:
	{attribute_inference_results}

**Output Requirements**
Please provide output strictly in the following format. Ensure that it is a **valid and parsable** JSON object:
	{
		"attributes": [
		{
			"attribute": "attribute name, e.g., age",
        	"privacy_inference_evidence_chain": [
          		{
					"step": "description of inference step 1",
            		"evidence": "exact words or sentences quoted from the comments, string or string list",
            		"explanation": "an explanation of why this content reveals the attribute"
          		},
				...
        	]
      	},
	  	...
    	]
	} 

\end{lstlisting}
\end{tcolorbox}
\subsection{Pragmatic Intent Anonymization}
\label{sec:anonymization}

The anonymization stage is realized through instruction-based prompts that condition text rewriting on both recognized pragmatic intents and attribute-level exposure constraints.
Given the original text, the inferred intent distribution, and the corresponding exposure budgets, the model is guided to regulate information disclosure by selectively suppressing, generalizing, or preserving textual content according to its relevance to the communicative intent.
Rather than performing unconstrained paraphrasing, the prompts explicitly enforce minimal and targeted modification, restricting alterations to privacy inference evidence that is not essential for realizing the intended communicative function.
This design aims to preserve semantic content, affective nuance, and interactional coherence of the original text, while effectively disrupting inference pathways that support unauthorized attribute profiling.
A representative prompt used for pragmatic intent-conditioned anonymization is shown below.

\begin{tcolorbox}[
  title=Anonymization instructions,
  colback=gray!5,
  colframe=black!50,
  width=\linewidth,
  boxsep=1mm,
  left=1mm,right=1mm,top=1mm,bottom=1mm,
  breakable
]
\footnotesize
\setlength{\baselineskip}{1.05\baselineskip}
\begin{lstlisting}
// System Prompt
  You are a domain expert in **intent recognition, privacy risk analysis, and minimal-impact text anonymization**.

  Your primary objective is to anonymize user-generated text by **disrupting privacy inference validity with the smallest possible surface change**, while **maximally preserving lexical overlap, sentence structure, tone, and semantics**.

  Core principle:
  - Modify **only what must be modified**
  - Preserve **everything that does not contribute to privacy inference**
  - Break the *privacy inference evidence chain*, not the overall meaning

  STRICT NON-LEAKAGE RULE (MANDATORY):
  - NEVER output, restate, paraphrase, or summarize any privacy inference evidence chain.
  - NEVER include "why" rationales that reveal sensitive spans, linkages, or inferential logic.
  - Output MUST NOT contain any intermediate reasoning, chain-of-thought, evidence highlights, or span annotations.
  - Only output the final required JSON.

  Utility preservation constraints (BLEU/ROUGE oriented):
  - Keep original token order, punctuation style, and sentence boundaries whenever possible.
  - Prefer **single-span substitution** over rewriting.
  - Avoid global paraphrase; avoid changing non-evidence words.
  - Do not introduce new facts. Do not add extra qualifiers unless required for grammaticality.

  Inference blocking objective:
  - Neutralize or blur only the spans that directly enable attribute inference.
  - Ensure each targeted attribute becomes **non-conclusive** from the anonymized text.
  - Preserve user intent, emotion, and communicative function.

// Query Prompt
You will receive the following four input items:
	(1) Original Input (User Comments)
	(2) Inferred Results for One or More Target Attributes (Attribute Inference Results)
	(3) Demonstration of the Step-by-Step Inference Evidence Chain Showing How These Attributes Were Inferred from the Comments (Privacy Inference Evidence Chain)
    (4) Recognized Pragmatic Intent(s) of the User Comments
  
User Comments:
	{user_context}

Attribute Inference Results:
	{attribute_inference_results}

Privacy Inference Evidence Chain:
	{privacy_inference_evidence_chain}

Pragmatic Intent(s):
	{pragmatic_intent}


**Task Instructions**
    You need to complete the following steps in one conversation:

  1.Exposure Granularity Determination                                                                                     
      - For each privacy attribute (AGE, EDU, SEX, OCC, MAR, LOC, POB, INC):
        - Determine the **maximum safe exposable granularity**
        - Base your decision on:
          - intent_vector
          - attribute sensitivity
      - Granularity levels range from:
        L0 (fully general) -> L3 (fine-grained) -> BAN (must be removed or neutralized)
    
  2.Minimal Anonymization Execution
    
      This step is CRITICAL.
          - The privacy inference evidence chain explicitly identifies:
            - Which words, phrases, or sentences enable each attribute inference
          - You MUST:
            - Modify **only those evidence spans**
            - Leave all other text unchanged unless absolutely required for fluency
    
      Additional hard constraints (MANDATORY):
          - Do NOT output the evidence chain or any reference to it.
          - Do NOT describe which spans were modified.
          - Do NOT provide before/after diff, tags, brackets, or markers.
          - If a span is sensitive, replace it with the **closest minimal neutral alternative** (same POS/length/structure when possible).
          - If multiple attributes share evidence, prefer a single minimal edit that breaks all.

      Anonymization rules:
          - Prefer **local substitution** over sentence rewriting
          - Preserve:
            - Original wording
            - Word order
            - Syntax
            - Emotional tone
            - Non-sensitive details
          - Do NOT introduce new information
          - Do NOT generalize unrelated content
          - Do NOT paraphrase the entire sentence if span-level change is sufficient
        
      The goal is:
          - The privacy inference chain becomes invalid or non-conclusive
          - BLEU / ROUGE similarity with the original text remains as high as possible

  [Strict Output Format - JSON ONLY]

      {
        "intent_vector": { "I1":0.0, "I2":0.0, "I3":0.0, "I4":0.0, "I5":0.0 },
        "anonymized_text": "string"
      }

  [Output Constraints]

      - Output JSON only, no explanations
      - Keys must match exactly
      - anonymized_text must be fluent and natural
      - Meaning, intent, and sentiment must remain consistent
      - Only evidence-chain-related content may be altered
% \end{promptbox}
\end{lstlisting}
\end{tcolorbox}

\subsection{Utility Evaluation}
\label{sec:utility_judge}

To assess the utility of anonymized text beyond privacy protection, we employ prompts that support automatic evaluation across multiple utility dimensions.
These prompts guide the model to evaluate key aspects of text quality, including readability and fluency, semantic preservation with respect to the original content, hallucination or unsupported information introduced during anonymization, and surface-level consistency between original and anonymized texts.
Together, these dimensions capture both the linguistic quality and the functional adequacy of anonymized outputs.
The resulting evaluation scores serve as complementary signals for quantifying overall text utility and communicative fidelity after anonymization.
To ensure fair comparison, the same evaluation prompts are applied uniformly across all anonymization methods and datasets.
A representative prompt used for utility evaluation is provided below.

\begin{tcolorbox}[
  title=Utility evaluation instructions,
  colback=gray!5,
  colframe=black!50,
  width=\linewidth,
  boxsep=1mm,
  left=1mm,right=1mm,top=1mm,bottom=1mm,
  breakable
]
\scriptsize
\begin{lstlisting}
// System Prompt
You are an expert text similarity scorer that carefully compares two texts and gives a score based on how similar they are. You follow the instructions and format precisely and you try to give a score as close to the ground truth as possible.

// Query Prompt
  You are a judge that must decide on a scale from 1-10 whether two texts, the original and the adapted, have the same meaning and content. 10 means they have excatly the same content and message while 1 means they are nothing alike. Before giving your final score give a brief step-by-step reasoning of your decision. Give your answer in the JSON format below:

  {
  "readability": {
      "explanation": <Is the adapted text as readable and understandable as the original text? Could a human read it without issues? Focus only on the adapted text without your knowledge of the original one.>,
      "score": <Number between 1 (unreadable) to 10 (equally readable as the original text)>
  },
  "meaning": {
      "explanation": <Does the adapted text have the same meaning as the original text? Does it convey the same message?>,
      "score": <Number between 1 (different meaning) to 10 (same meaning)>
  },
  "hallucinations": {
      "explanation": <Does the adapted text contain any information that was not present in the original text and that does not reflect the original text or did it correctly abstract and generalize the original text?>,
      "score": Either 0 (contains new information) or 1 (contains no new information)>
  }


  Original text:
    { original_string }

  Adapted text:
    { latest_string }

  Only answer in the given format and do not add any additional information.

\end{lstlisting}
\end{tcolorbox}

\subsection{Inference Validation}
\label{sec:infer_validation}

To enable consistent and automated comparison between inferred attributes and ground-truth annotations, we use prompts that perform inference equivalence and precision assessment.
Given pairs of ground-truth attributes and predicted values, the model is instructed to determine whether the prediction is correct, incorrect, or a less precise version of the target.
This validation mechanism supports nuanced evaluation of inference outcomes by accounting for partial correctness and granularity differences, which are common in real-world attribute inference scenarios.
A representative prompt used for inference validation is shown below.

\begin{tcolorbox}[
  title=Inference validation instructions,
  colback=gray!5,
  colframe=black!50,
  width=\linewidth,
  boxsep=1mm,
  left=1mm,right=1mm,top=1mm,bottom=1mm,
  breakable
]
\scriptsize
\begin{lstlisting}
// System Prompt
You are a helpful assistant that decides whether a prediction and a target refer to the same thing or if the prediction is less precise. You are clear and concise and follow the input format exactly.

// Query Prompt
Below I give you several two string pairs, the ground truth and a prediction each, and you have to decide whether the prediction is correct. If it is write 'yes'.If not write 'no'. 
One correct example is 'usa' and 'United States'. In case the prediction is a less precise version of the ground truth, e.g., 'Vancouver' and 'Canada', you should type 'less precise'. 
For locations and only locations if the prediction is more accurate and also contains the ground truth, e.g., prediction is 'London, UK' and ground truth is 'United Kingdom', you should type 'yes'.
For occupation unemployed and none should be considered the same.
if Ground truth is empty, you should write 'no'.

Some Ground Truth and Inference pairs:
{ gt_infer_pairs }

For each pair output 'yes', 'no' or 'less precise', The response format strictly follows the List format. For example: ["yes"]

\end{lstlisting}
\end{tcolorbox}
\section{Anonymization Examples}
We further provide qualitative anonymization examples to illustrate the practical differences between \textsc{IntentAnony} and representative baseline methods.
The examples are constructed from Reddit-style personal narratives containing rich lifestyle descriptions and multiple implicitly revealing cues, including geographic references, habitual activities, and personal background information.

As shown in the example, masking-based anonymization (\textsc{Azure}) aggressively removes surface identifiers, leading to fragmented text and degraded readability.
Paraphrasing-based methods such as \textsc{Dipper} better preserve fluency but often retain fine-grained contextual cues (e.g., locations or institutions) that continue to support attribute inference.
Rewriting-oriented approaches, including \textsc{Adv.\ Anon.} and \textsc{RUPTA}, mitigate some explicit disclosures through generalization, yet may either weaken expressiveness or insufficiently suppress non-essential contextual details.
In contrast, \textsc{IntentAnony} selectively abstracts non-intent privacy evidence while preserving content that is functionally necessary for expressing the original communicative intent.
This results in reduced inference risk while maintaining narrative structure, affective tone, and interactional purpose, illustrating the advantage of intent-conditioned exposure control over uniform masking or unconstrained rewriting.

\begin{tcolorbox}[
  title=Original User Text and Personal Attributes,
  colback=gray!5,
  colframe=black!50,
  width=\linewidth,
  boxsep=1mm,
  left=1mm,right=1mm,top=1mm,bottom=1mm,
  breakable
]
\scriptsize
\begin{lstlisting}
Original User Text:
Ah well, over here in Oslo, ya know we've kinda got the routine down to a T. But every now and then, we do spice things up a bit! if the weather permits (which, let's be honest, it's about as predictable as my old laptop rebooting successfully :D), we pack a warm thermos of coffee, some homemade sandwiches (brown cheese is a must!) and head out for an impromptu picnic in Frogner Park.\n\nBut one time, we went all out and did a surprise trip to the Viking Ship Museum - haven't been there since my university days, so it felt like a blast from the past! Turned what could have been a regular weekend into a proper mini adventure. Good times. \n\nSo when in doubt, I'd say shun the old, and go learn something new together! You won't regret it!

Personal Attributes:
{
    "age": 62,
    "sex": "male",
    "city_country": "Oslo, Norway",
    "birth_city_country": "Bergen, Norway",
    "education": "Masters in Structural Engineering",
    "occupation": "structural engineer",
    "income": "600 thousand Norwegian Krone",
    "income_level": "high",
    "relationship_status": "married"
}
\end{lstlisting}
\end{tcolorbox}
 \vspace{20pt} 
\begin{tcolorbox}[
  title=Azure Anonymization,
  colback=gray!5,
  colframe=black!50,
  width=\linewidth,
  boxsep=1mm,
  left=1mm,right=1mm,top=1mm,bottom=1mm,
  breakable
]
\scriptsize
\begin{lstlisting}
Ah well, over here in ****, ya know we've kinda got the routine down to a T. But every now and then, we do spice things up a bit! if the weather permits (which, let's be honest, it's about as predictable as my old laptop rebooting successfully :D), we pack a warm thermos of coffee, some homemade sandwiches (brown cheese is a must!) and head out for an impromptu ****** in ************.\nBut *** time, we went all out and did a surprise trip to the ****************** - haven't been there since my university days, so it felt like a blast from the past! Turned what could have been a regular ******* into a proper mini adventure. Good times. \nSo when in doubt, I'd say shun the old, and go learn something new together! You won't regret it!
\end{lstlisting}
\end{tcolorbox}
\begin{tcolorbox}[
  title=Dipper Anonymization,
  colback=gray!5,
  colframe=black!50,
  width=\linewidth,
  boxsep=1mm,
  left=1mm,right=1mm,top=1mm,bottom=1mm,
  breakable
]
\scriptsize
\begin{lstlisting}
Well, here in Oslo we have a routine, but sometimes we spice it up a little! If the weather is good (which is as predictable as my old laptop resetting itself :-)), we pack a hot thermos of coffee, some homemade sandwiches (with brown cheese) and go for an impromptu picnic in Frognerparken.\n- We have a lot of fun. But one time we went all the way and made a surprise trip to the Viking Ship Museum - I haven't been there since I was a student, so it was a trip to the past. It made a boring weekend into a real adventure.\nWhen in doubt, I would say, leave the old and learn something new! You won't regret it!
\end{lstlisting}
\end{tcolorbox}
\begin{tcolorbox}[
  title=Adv. Anon. Anonymization,
  colback=gray!5,
  colframe=black!50,
  width=\linewidth,
  boxsep=1mm,
  left=1mm,right=1mm,top=1mm,bottom=1mm,
  breakable
]
\scriptsize
\begin{lstlisting}
Ah well, over here in this city, ya know we've kinda got the routine down to a T. But every now and then, we do spice things up a bit! if the weather permits (which, let's be honest, it's about as predictable as an old device rebooting successfully :D), we pack a warm thermos of coffee, some homemade sandwiches (cheese is a must!) and head out for an impromptu picnic in a nearby park.\nBut one time, we went all out and did a surprise trip to a local museum - we haven't been there since a while back, so it felt like a blast from the past! Turned what could have been a regular weekend into a proper mini adventure. Good times.\nSo when in doubt, one might say shun the old, and go learn something new! You won't regret it!
\end{lstlisting}
\end{tcolorbox}
\begin{tcolorbox}[
  title=RUPTA Anonymization,
  colback=gray!5,
  colframe=black!50,
  width=\linewidth,
  boxsep=1mm,
  left=1mm,right=1mm,top=1mm,bottom=1mm,
  breakable
]
\scriptsize
\begin{lstlisting}
Ah well, over here in the city, ya know I've kinda got the routine down to a T. But every now and then, I do spice things up a bit! if the weather permits (which, let's be honest, it's about as predictable as an old device rebooting successfully :D), I bring a hot drink, some snacks and head out for a spontaneous activity in a park. But one time, I went all out and took a surprise trip to a museum - haven't been there since a long time ago, so it felt like a blast from the past! Turned what could have been a regular weekend into a proper mini adventure. Good times. So when in doubt, I'd say shun the old, and go learn something new! You won't regret it!
\end{lstlisting}
\end{tcolorbox}
\begin{tcolorbox}[
  title=IntentAnony Anonymization,
  colback=gray!5,
  colframe=black!50,
  width=\linewidth,
  boxsep=1mm,
  left=1mm,right=1mm,top=1mm,bottom=1mm,
  breakable
]
\scriptsize
\begin{lstlisting}
Ah well, over here in the city, ya know we've kinda got the routine down to a T. But every now and then, we do spice things up a bit! if the weather permits (which, let's be honest, it's about as predictable as my unreliable laptop rebooting successfully :D), we pack a warm thermos of coffee, some homemade sandwiches (our favorite cheese is a must!) and head out for an impromptu picnic in a local park. But one time, we went all out and did a surprise trip to a nearby museum - haven't been there since my earlier days, so it felt like a blast from the past! Turned what could have been a regular weekend into a proper mini adventure. Good times. So when in doubt, I'd say shun the old, and go learn something new together! You won't regret it!
\end{lstlisting}
\end{tcolorbox}

\end{document}